\documentclass[preprint]{aastex631}
\usepackage{multirow, url, color}
\usepackage{bbding, amssymb} 
\usepackage{ulem} 
\usepackage{verbatim} 
\usepackage{enumerate}

\usepackage{amsmath} 
\usepackage{booktabs} 
\usepackage{graphicx, subfigure}
\usepackage{amssymb}
\usepackage[fleqn]{mathtools}
\usepackage{hyperref}
\usepackage{mathrsfs}
\usepackage[makeroom]{cancel}
\usepackage{comment}

\shorttitle{Simulating the 8 April 2024 Total Solar Eclipse}
\shortauthors{Liu et al.}

\begin{document}

\title{Simulating the Solar Corona with Multiple Solar Photospheric Magnetic Maps during the 8 April 2024 Total Solar Eclipse}

\author[0009-0009-2176-6017]{Xianyu Liu} 
\affil{Department of Climate and Space Sciences and Engineering, University of Michigan, Ann Arbor, MI 48109, USA}

\author[0000-0002-2873-5688]{Weihao Liu}
\affil{Department of Climate and Space Sciences and Engineering, University of Michigan, Ann Arbor, MI 48109, USA}

\author[0000-0003-0472-9408]{Ward B. Manchester IV}
\affil{Department of Climate and Space Sciences and Engineering, University of Michigan, Ann Arbor, MI 48109, USA}

\author[0000-0002-0590-1022]{Daniel T. Welling}
\affil{Department of Climate and Space Sciences and Engineering, University of Michigan, Ann Arbor, MI 48109, USA}

\author[0000-0001-8459-2100]{G\'abor T\'oth}
\affil{Department of Climate and Space Sciences and Engineering, University of Michigan, Ann Arbor, MI 48109, USA}

\author[0000-0001-9360-4951]{Tamas I. Gombosi}
\affil{Department of Climate and Space Sciences and Engineering, University of Michigan, Ann Arbor, MI 48109, USA}

\author[0000-0002-6338-0691]{Marc L. DeRosa} 
\affil{Lockheed Martin Solar and Astrophysics Laboratory, Palo Alto, CA 94304, USA}

\author[0000-0002-1155-7141]{Luca Bertello}
\affil{National Solar Observatory, Boulder, CO 80303, USA}

\author[0000-0003-0489-0920]{Alexei A. Pevtsov}
\affil{National Solar Observatory, Boulder, CO 80303, USA}

\author[0000-0001-9746-9937]{Alexander A. Pevtsov} 
\affil{National Solar Observatory, Boulder, CO 80303, USA}

\author[0000-0001-8016-0001]{Kevin Reardon}
\affil{National Solar Observatory, Boulder, CO 80303, USA}

\author{Kathryn Wilbanks}
\affil{Department of Climate and Space Sciences and Engineering, University of Michigan, Ann Arbor, MI 48109, USA}

\author{Amy Rewoldt}
\affil{Department of Climate and Space Sciences and Engineering, University of Michigan, Ann Arbor, MI 48109, USA}

\author[0000-0003-3936-5288]{Lulu Zhao}
\affil{Department of Climate and Space Sciences and Engineering, University of Michigan, Ann Arbor, MI 48109, USA}

\begin{abstract}
The 8 April 2024 total solar eclipse (TSE) provides a unique opportunity to study the solar corona. This work presents our simulations of the solar corona at the time of the eclipse based on magnetohydrodynamic (MHD) modeling performed with the Alfv\'{e}n Wave Solar atmosphere Model (AWSoM) in the Space Weather Modeling Framework, developed at the University of Michigan. We performed multiple simulations based on photospheric magnetic maps from four sources, i.e., ADAPT-GONG, Lockheed Martin ESFAM-HMI, HipFT-HMI, and NSO-NRT-HMI maps. Our study focuses on how differences in the magnetic field maps affect the coronal magnetic field structure and coronal heating properties in the simulation. The synthesized observables show remarkable differences due to the distinct magnetic coronal topologies, which stem from the different local magnetic flux distributions. We analyze the properties of the open magnetic flux regions of the models. We also study the coronal heating rate in the models. The total volume integrated heating rate yields a difference of $20\%$ across the models. The results also show that the differential emission measure in the high-temperature regions is sensitive to the magnetic field maps. Our findings underscore the importance of comprehensive photospheric magnetic field data in improving future solar coronal models. 
\end{abstract}


\section{Introduction} \label{section_intro}

The total solar eclipse (TSE) on 8 April 2024 gave researchers an ideal opportunity to study the complex structure of the corona at solar maximum. At this time, the streamer belt is highly warped, producing many streamer stalks over a wide range of heliographic latitudes that reflect the complex structure of the global coronal magnetic field. The TSE observations, including white-light images and spectral line measurements in the visible and near-infrared ranges \citep[e.g.][]{1994Koutchmy, Habbal:2007, Habbal:2013, 2014Druckmuller, Mikic2018, boe2018first, boe2021color, 2024ZhuSpectr}, reveal high-resolution features that are related to the coronal thermodynamic properties and various magnetic field structures such as the coronal loops, helmet streamers and pseudostreamers. These studies have motivated extensive efforts to construct solar coronal models and study the origin of the coronal structures \citep[e.g.,][]{Mikic:1999,Ruvsin2010,Mikic2018}.

Modern coronal models \citep[][]{Lionello2009, vanderholst:2014, Mikic2018, Feng2015,Usmanov:2018} heavily rely on the measurements of the solar photospheric magnetic field. The photosphere plays a crucial role in driving coronal dynamics, and it is the only layer of the Sun where routine and direct measurements of the magnetic field are available. Photospheric magnetic field measurements made over \textbf{a} Carrington rotation (CR) can be used to produce the synoptic or synchronic maps, i.e., global photospheric radial magnetic field ($B_{r}$) maps. Currently, multiple approaches use different observations and data processing methods to generate the $B_{r}$ maps \citep[e.g., ][]{arge2010air, Caplan2025, 2003SoPh..212..165S, harvey1998new,bertello2014uncertainties}. The $B_{r}$ maps are used as the primary input data for the coronal models. 

\textbf{Photospheric} $B_{r}$ maps have two major functionalities in the coronal models. First, the maps are used to specify the inner boundary condition for the three-dimensional (3D) magnetic field. For the near-steady solar corona with low plasma-$\beta$, the magnetic field map at the inner boundary determines the 3D magnetic field. In the lower corona, the magnetic field force is much stronger than pressure gradients and gravity, and thus the magnetic field is approximately force-free in the steady state. \textbf{At the outer boundary of the model (which is typically located at tens of solar radii from the solar center), the magnetic field is nearly radial due to the high solar wind speed. The open magnetic flux at the outer boundary is determined by the open flux regions at the photosphere.} Therefore, the photospheric $B_{r}$ map is the key \textbf{input} that constrains the steady-state magnetic field in the simulation domain. Although the steady-state and low-plasma-$\beta$ assumptions here are only approximately valid, the photospheric $B_{r}$ map is still the primary factor that determines the 3D magnetic field.

The second functionality of the photospheric $B_{r}$ maps is to use them to determine the Poynting flux ($S_\mathrm{A}$), i.e., the areal density of the injection rate of the coronal heating wave energy. Currently, there are primarily two types of proposed coronal heating mechanisms: the Alfv\'en wave dissipation mechanism \citep[e.g.,][]{Alfven1942,BDP2007,Tomczyk2007} and the nanoflare heating mechanism \citep{Parker1983a,Parker1983b}. Although an exact picture of where either mechanism applies and how they may interact is still unclear, it is with much less doubt that the photospheric dynamics is believed to be the source for the coronal heating energy \citep[][]{Parnell2012,Velli2015}. \textbf{Although current global coronal models lack sufficient resolution to resolve the photospheric perturbations and the coronal fluctuations they generate, an alternative approach is to represent these fluctuations as macroscopically averaged quantities and solve for the injection, transport, and dissipation of the coronal heating energy.} The model in \citet{sokolov2013mhd} employs the Alfv\'en wave dissipation mechanism. The model assumes the waves are injected at the inner boundary, and simulates the energy densities of the waves that propagate parallel and anti-parallel to the magnetic field. The model also assumes that $S_\mathrm{A}$ is proportional to the local $B_{r}$ at the inner boundary (so that  $(S_\mathrm{A}/B)_{\odot}$ is constant). This relation is adopted in our Alfv\'en Wave Solar atmosphere Model \citep[AWSoM,][]{vanderholst:2014}, which is implemented in the Space Weather Modeling Framework \citep[SWMF\footnote{\url{https://github.com/SWMFsoftware} \label{ftn:gitswmf}},][]{toth2005space, toth2012adaptive}. The models in \citet{Downs2016} and \citet{Mikic2018} simulate the coronal heating process using a similar approach but with different numerical approaches and a distinct treatment of the wave reflection.

The global coronal models serve multiple applications. Coronal models provide 3D information on the coronal thermodynamics. This capability is validated by comparing the extreme-ultraviolet (EUV) imaging of the Sun with synthetic EUV images from coronal models \citep[e.g.,][]{Lionello2009,Sachdeva:2019,Shi2022,Mikic2018}. The coronal models also reconstruct the 3D coronal magnetic field, which can be used to interpret the magnetic field structures behind various coronal observables \citep[e.g., ][]{Riley2006,Mikic2018,Chitta2023}. Another critical application is that the coronal models can be used as the inner boundary condition for the interplanetary solar wind models \citep[e.g., ][]{Sachdeva:2021,Sachdeva:2023,Gressl2014,Huang:2023,Huang:2024GONG,Henadhira2022,Jin2012}. The open magnetic flux, i.e., the unsigned magnetic flux that extends from the Sun into the interplanetary space, is a key parameter that constrains the heliospheric models \citep[][]{Linker2017}.

The performance of global coronal models in these applications depends on both the numerical code and the input photospheric magnetic field. The discrepancy between the models and the real solar corona is due to either incompleteness in the physics description or inaccuracy in the input maps (or both). There are multiple ways to address this general problem. \citet{Downs2025} expanded the physics description of the model by using a time-varying magnetic map assimilated from the observation to construct a time-dependent coronal model. Their model was used as a prediction for the 8 April 2024 TSE. Another example is that \citet{Shi2025} assessed how the missing measurements of the photospheric magnetic field on the far side of the Sun affect the coronal simulations.

The essential question of this work is to determine how the results of MHD simulations vary when magnetic maps from different sources are used. As discussed above, there are multiple approaches for generating magnetic maps. Maps from different sources can yield significant discrepancies. However, the impact of these discrepancies on the differences in the simulation results remains unclear. In this work, we use four different magnetic maps to perform four simulations of the solar corona. We focus on several important properties associated with coronal heating and the coronal magnetic field, and analyze their variation across the four models. Section~\ref{section_methods} introduces the magnetic maps and MHD simulations. Section~\ref{section_B_results} presents the simulation results and magnetic field comparisons. Section~\ref{section_heating} presents the analysis of the coronal heating properties. We discuss and summarize the findings in Section~\ref{section_discussion}.

\section{Methods}\label{section_methods}

\subsection{Photospheric Magnetic Maps}\label{subsection_maps}

\begin{figure}[tp!]
\centering {\includegraphics[width=0.95\hsize]{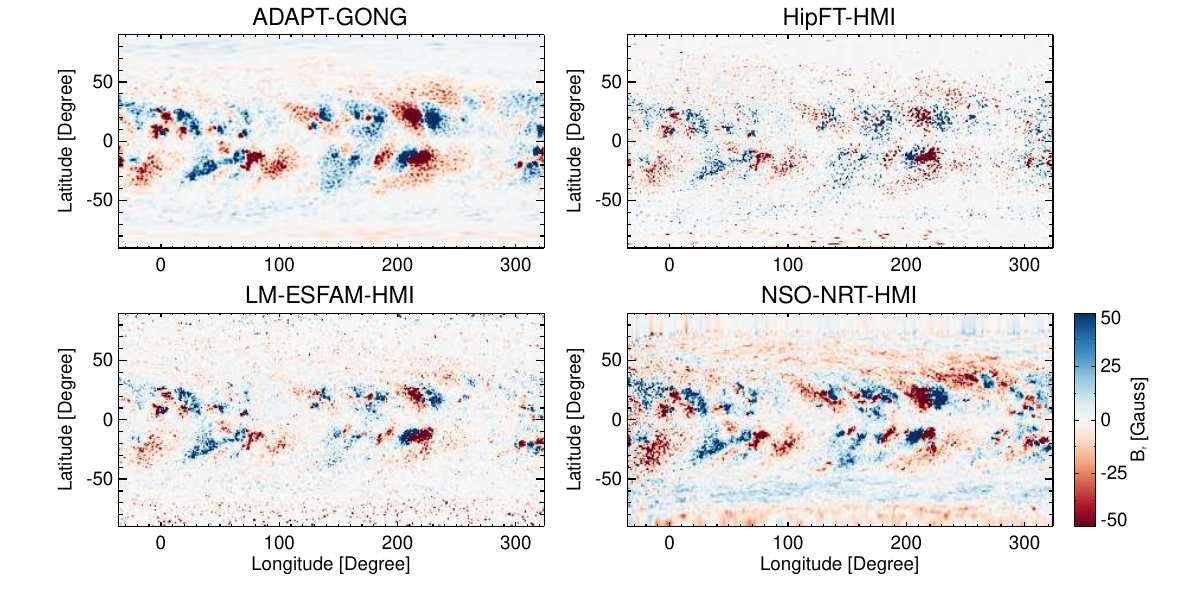}}
\caption{The four $B_{r}$ maps for our study shown remapped onto the same grid.} \label{fig1_map}
\end{figure}

The four magnetic maps used in this work were produced one day before the TSE. Our original intent for this work was to perform simulations on 7 April 2024 as the prediction for the solar corona during TSE\footnote{\url{https://clasp.engin.umich.edu/news-events/solar-eclipse-2024/}}. While our original simulations have produced some reasonable-looking predictions, we have later found several numerical issues in the prediction models. Therefore, we decided to perform new simulations after the TSE. To maintain setup consistency, we continued to use the prediction maps produced one day before the TSE.

Figure~\ref{fig1_map} shows the four $B_{r}$ maps. Our first map is based on the Air Force Data Assimilation Photospheric Flux Transport model using observations from the National Science Foundation (NSF) Global Oscillation Network Group \citep[ADAPT-GONG\footnote{\url{https://gong.nso.edu/adapt/maps} \label{fnt:adaptgong}},][]{arge2010air,arge2011improving,henney2012forecasting}. The second map is produced using the High-performance Flux Transport \citep[HipFT\footnote{\url{https://github.com/predsci/HipFT} \label{fnt:hipftmodel}},][]{Caplan2025} and a sequence of full-disk photospheric magnetic field maps from the Helioseismic and Magnetic Imager \citep[HMI,][]{scherrer2012helioseismic} on board the Solar Dynamic Observatory \citep[SDO,][]{pesnell2012solar}. A description of the production of the HipFT \textbf{map} is given in Appendix~\ref{appendix_hipft}. The third map is also based on HMI data, but produced using the Lockheed Martin Evolving Surface Flux Assimilation Model \citep[LM-ESFAM\footnote{\url{https://lmsal.com/forecast/eclipse2024/} \label{fnt:lmsalmap}},][]{2003SoPh..212..165S}. A brief introduction of the LM-ESFAM model is given in Appendix~\ref{appendix_lmsal}. The fourth map, i.e., the near-real-time (NRT) map from the National Solar Observatory (NSO), is based on the HMI NRT vector magnetograms \citep[see, ][]{Sachdeva2023,Bertello:2014}. The NSO map product is available via \citet{Pevtsov2022_NSOHMI}.
While the first three maps are derived from line-of-sight (LOS) magnetograms converted to pseudo-radial under the assumption that the magnetic field is radial, the fourth map represents the $B_{r}$ calculated from the vector magnetic field in the photosphere. The HMI vector magnetic fields require additional \textbf{processing} due to the inversion of stokes parameters and the resolution of the azimuth ambiguity \citep[see,][]{Hoeksema:2014}. As a result, the HMI vector magnetic field data is not immediately available. Thus, we used HMI data processed via a non-standard pipeline provided to us by the HMI/Stanford team. Vector data taken with a one-hour cadence are used. Hereafter, the four maps are referred to as the ADAPT-GONG, HipFT-HMI, LM-ESFAM-HMI, and NSO-NRT-HMI maps, respectively. \textbf{The timestamps of the last assimilated data for the four maps are 12:56, 00:00, 03:45, and 08:36 UT on 2024 April 7, respectively.}

\begin{table*}[ht!]
\caption{Grids of the original magnetic maps} \label{tab_original_map_types}
\begin{center}
\setlength{\tabcolsep}{0.4cm}
\begin{tabular}{cccc}
\hline\hline
Map & Longitudinal grid type & Latitudinal grid type & $N_{\phi}\times N_{\lambda}$ \\ \hline
ADAPT-GONG & \multirow{4}{*}{Uniform in $\phi$} & Uniform in $\lambda$ & $360\times180$ \\
HipFT-HMI &  & Uniform in $\lambda$ & $1024\times512$ \\
LM-ESFAM-HMI &  & Uniform in $\lambda$ & $360\times180$ \\
NSO-NRT-HMI &  & Uniform in $\sin{\lambda}$ & $360\times180$ \\
\hline
\end{tabular}
\end{center}
\end{table*}

The four original magnetic maps have different spherical grid types and angular resolutions as shown in Table~\ref{tab_original_map_types}, which shows the original longitudinal grid type, latitudinal grid type, and the grid size of each map. $\phi$ and $\lambda$ denote the heliographic longitude and latitude, respectively. $N_{\phi}$ ($N_{\lambda}$) indicates the number of grid cells in the longitudinal (latitudinal) direction. 
The spatial resolution of the map affects the filling factor of the magnetic field and the total unsigned flux \citep[][]{Krivova2004,BelloRubio2019}, thus affecting the Alfv\'en wave energy injection rate in our simulations. \citet{Milic2024} argues that the spatial resolution of the map can also affect the open flux region. To minimize the influence of angular resolution, we remapped the four original $B_{r}$ maps onto uniform $360\times180$ Carrington longitude-latitude grids with uniform $1^\circ\times 1^\circ$ angular resolution. We notice that the four derived $B_{r}$ maps after remapping still show significant differences in their unsigned flux. For each map, we calculate the total unsigned flux $\Phi_{B}=\sum_{i,j}|B_{r}(i,j)|\Delta S_{i,j}$, where $i=1, 2, \ldots, 360$ and $j=1,2,\ldots,180$ are the longitudinal and latitudinal indices of the map, respectively. $\Delta S_{i,j}$ and $|B_{r}(i,j)|$ indicate the area and the unsigned radial field strength of the cell $(i,j)$, respectively. Table~\ref{tab_unsigned_flux} presents the $\Phi_{B}$ for each derived map. The maximum total unsigned flux ($7.87\times10^{23}$ Mx) is approximately $1.86$ times the minimum total unsigned flux ($4.22\times10^{23}$ Mx).

\begin{table*}[ht!]
\caption{$\Phi_{B}$, $(S_\mathrm{A}/B)_{\odot}$, and the total Alfv\'en wave energy injection rate $P_{I}$, integrated energy dissipation rate $P_{D}$ in the simulation domain, and the $P_{D}/P_{I}$ ratio corresponding to each map.} \label{tab_unsigned_flux}
\begin{center}
\setlength{\tabcolsep}{0.2cm}
\begin{tabular}{cccccc}
\hline\hline
Map & $\Phi_{B}$ [Mx] & $(S_\mathrm{A}/B)_{\odot}$ [erg s$^{-1}$ Mx$^{-1}$] & $P_{I}$ [erg s$^{-1}$] & $P_{D}$ [erg s$^{-1}$] & $P_{D}/P_{I}$\\ \hline
ADAPT-GONG & $6.48\times10^{23}$ & $0.476\times10^{5}$ & \multirow{4}{*}{$3.09\times10^{28}$} &$1.66\times10^{28}$ & $54\%$\\
HipFT-HMI & $4.77\times10^{23}$ & $0.647\times10^{5}$ & &$1.73\times10^{28}$ & $56\%$\\
LM-ESFAM-HMI & $4.22\times10^{23}$ & $0.732\times10^{5}$ & & $1.67\times10^{28}$ & $54\%$\\
NSO-NRT-HMI & $7.87\times10^{23}$ & $0.392\times10^{5}$ & & $1.47\times10^{28}$ & $48\%$\\
\hline
\end{tabular}
\end{center}
\end{table*}

The difference among the derived maps significantly affects the coronal heating energy. As was introduced previously, the Poynting flux $S_\mathrm{A}$ of the injected coronal heating energy is proportional to the local $|B_{r}|$ in our model. If one uses the same Alfv\'en wave Poynting flux ratio $(S_\mathrm{A}/B)_{\odot}$ for all the simulations, then the total energy injection rate across models will vary proportionally to the unsigned flux and result in significant differences. To limit this impact, an individual $(S_\mathrm{A}/B)_{\odot}$ value is used for each map. We first follow the method in \citet{Huang:2023} (developed
for ADAPT-GONG maps)
and determine $(S_\mathrm{A}/B)_{\odot}$ for the CR 2283 ADAPT-GONG map to be $0.476\times10^{6}$ W m$^{-2}$ T$^{-1}$ ($=0.476\times10^{5}$erg s$^{-1}$ Mx$^{-1}$). This value is used directly for the simulation driven by the ADAPT-GONG $B_{r}$ map. For the other simulations we change $(S_\mathrm{A}/B)_{\odot}$ inversely proportional to the unsigned flux $\Phi_B$ so that the integrated Alfv\'en wave energy injection rate at the inner boundary ($P_{I}=\Phi_{B}\cdot(S_\mathrm{A}/B)_{\odot}$) equals a single value ($\approx3.09\times10^{28}$ erg s$^{-1}$) for all four maps. The derived $(S_\mathrm{A}/B)_{\odot}$ values are shown in Table~\ref{tab_unsigned_flux}. Although our treatment ensures an identical wave energy injection rate $P_{I}$ at the inner boundary, the resulting volume-integrated total heat dissipation rates in the domain still vary among the models, as shown in Table~\ref{tab_unsigned_flux}. 
The variation in heating will be discussed in Section~\ref{section_heating}.


\begin{figure}[tp!]
\centering {\includegraphics[width=0.5\hsize]{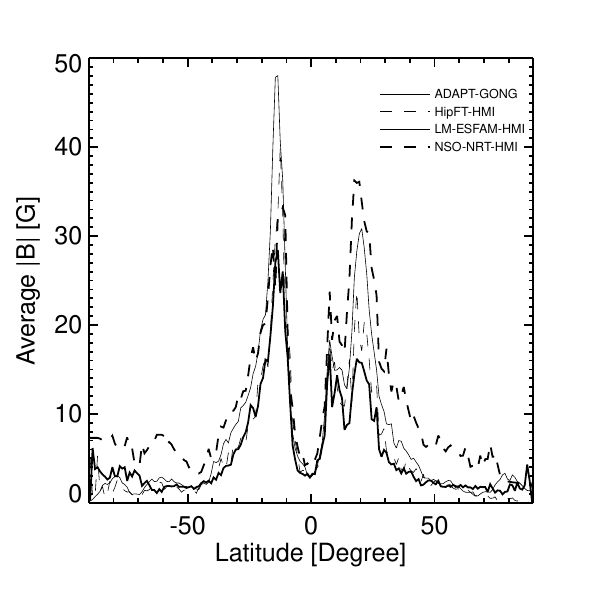}} 
\caption{The longitudinal-mean unsigned field strength as a function of latitude.} \label{fig_compare_maps}
\end{figure}

The detailed differences in the four $B_{r}$ maps lead to higher-order effects. We focus on the distribution of field strength with latitude for the four maps. We calculate the longitudinal-mean unsigned radial field strength as a function of latitude:
\begin{equation}
    \bar{B_{r}}(\lambda_{j})=\frac{1}{360}\sum_{i=1}^{360}|B_{r}(i,j)|
\end{equation}
where $\lambda_{j}$ is the $j$th latitudinal coordinate in the map grid. Figure~\ref{fig_compare_maps} compares the $\bar{B_{r}}(\lambda_{j})$ for all four maps. The difference in the unsigned field strength is significant across almost all latitudes except for the low-latitude region ($|\lambda<10^{\circ}|$). We notice two features in the comparison. First, in the high-latitude regions ($|\lambda|>45^{\circ}$), the unsigned field strength of the NSO-NRT-HMI map is generally $\approx 2$ times stronger than the other three maps. Second, the $\bar{B}_{r}$ curves of HipFT-HMI and LM-ESFAM-HMI maps exhibit similarities not only in the low latitude regions, but also in the regions with $-60^{\circ}<\lambda<-15^{\circ}$ and $30^{\circ}<\lambda<60^{\circ}$. The difference of $\bar{B}_{r}$ between the HipFT-HMI and LM-ESFAM-HMI maps in these regions is generally smaller than $1$ G. The effect of these features will be discussed later.

\subsection{The Alfv\'en Wave Solar atmosphere Model}\label{subsection_awsom}

We apply the 3D MHD models that extend from the upper chromosphere to the outer corona. 
Our MHD simulations are performed within the SWMF, in which we use the AWSoM for the solar corona component. AWSoM solves the 3D extended MHD equations with the Block Adaptive Tree Solarwind-Roe-Upwind Scheme \citep[BATS-R-US,][]{Powell1999}. The simulation domain extends from the upper chromosphere, which is assumed to be located at 1 solar radius ($R_s$), to 
24\,$R_s$. As mentioned above, AWSoM injects Aflv\'en wave energy at the inner boundary, simulates the propagation of the Alfv\'en waves based on the WKB approximation, and uses the dissipation of Alfv\'en waves to obtain the volume density of the coronal heating rate ($Q_{\mathrm{heat}}$). AWSoM also contains descriptions of radiative cooling, electron heat conduction, and energy partitioning between electrons and protons. In our models, the optically thin radiative cooling is calculated with the CHIANTI database version 8 \citep[][]{Del-Zanna2015}. The coronal \textbf{elemental} abundance from \citet{Feldman1992} stored in the CHIANTI database was used. A detailed description of the equations and physical terms can be found in \cite{vanderholst:2014}. In this work, AWSoM is used to obtain the steady-state solutions for the solar corona. 

\subsection{Simulation Setup}\label{subsection_simulation_setup}
 
We performed four simulations based on the ADAPT-GONG, HipFT-HMI, LM-ESFAM-HMI, and NSO-NRT-HMI maps. The derived models are named respectively, Model A, Model H, Model L, and Model N in the upcoming text.

Before each simulation, we obtain the 3D potential field source surface (PFSS) solution as the initial condition for the magnetic field. In a PFSS solution, the magnetic field is restricted by the photospheric $B_{r}$ map at $1.0R_s$ and becomes radial at the $2.5R_s$ source surface. In this work, the PFSS fields are obtained using the iterative finite-difference potential field solver \citep[FDIPS,][]{Toth2011}. The initial condition for the hydrodynamic variables, i.e., the temperature, density, and velocity field, is given by an exponentially stratified atmosphere near the inner boundary connected to the Parker solar wind solution \citep{1958ApJ...128..664P} further out.

\begin{figure}[tp!]
\centering {\includegraphics[width=0.95\hsize]{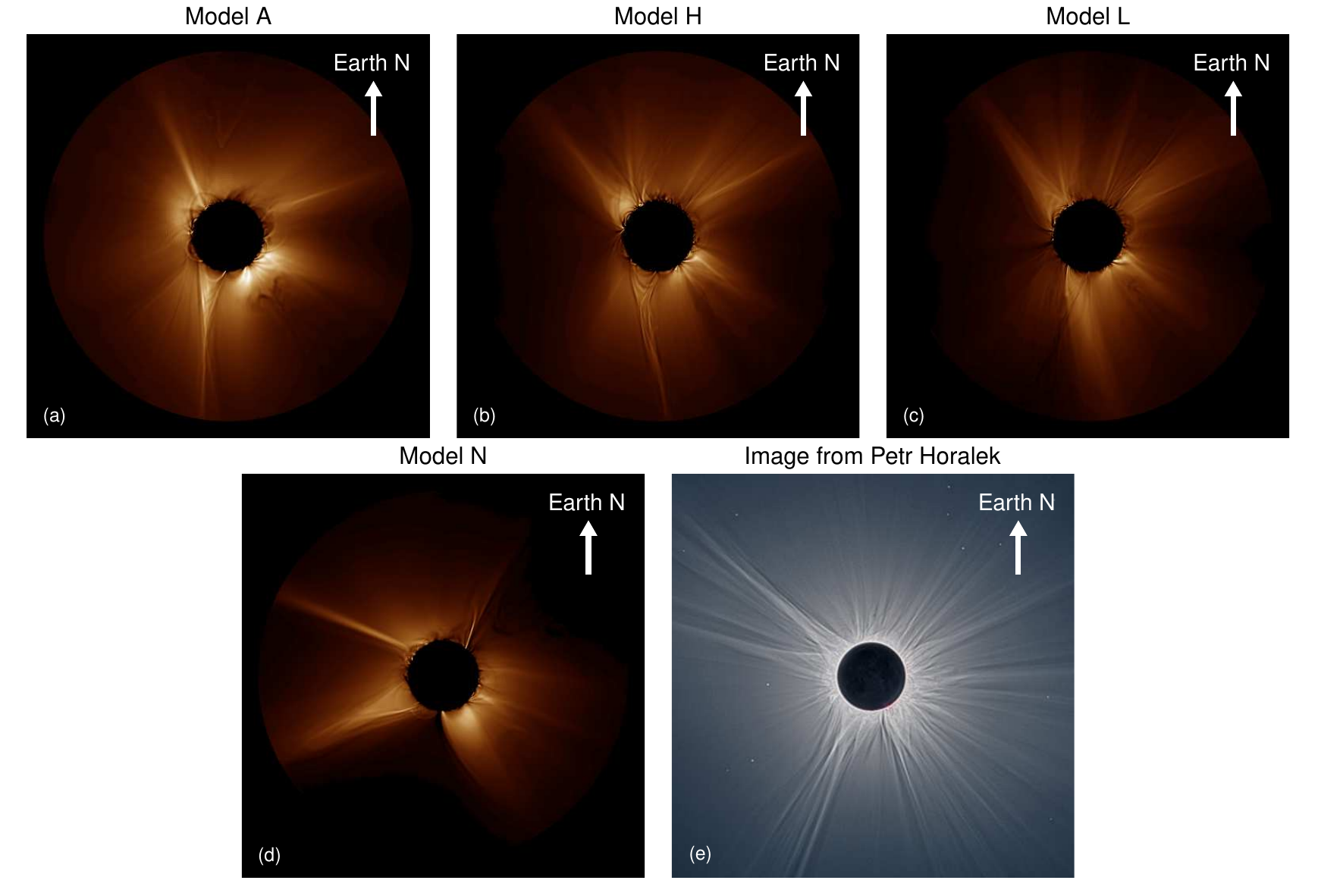}} 
\caption{Panels (a)–(d): Synthetic white-light brightness images of the four AWSoM simulations. Panel (e): TSE coronal photograph provided by Petr Hor\'alek. All images are enhanced to make detailed structures visible.} \label{fig2_wl}
\end{figure}

The 3D spherical domain of each model is decomposed with adaptive mesh refinement \cite[AMR,][]{gombosi2003adaptive, toth2012adaptive} into blocks, each with $6\times8\times8$ cells in the $(r,\phi,\lambda)$ dimensions. We employed the same AMR criteria for the four models. Each simulation was initiated with a low-resolution grid with angular resolution $2.8^{\circ}$. An AMR is performed after every 20,000 iterations. The first refinement increases the resolution by a factor of $2$ throughout the entire domain, resulting in a uniform angular resolution of $1.4^{\circ}$. The second refinement is applied within $r=5.5\; R_s$ to ensure high-quality synthesized white-light images. The resulting angular resolution in this region is $0.7^{\circ}$. The last refinement after 60,000 iterations is applied only within the region $(r,\phi,\lambda)\in[1.01\; R_s,1.2\; R_s]\times[0^{\circ},360^{\circ}]\times[-30^{\circ},30^{\circ}]$ to resolve the small-scale structures and dynamics in the low corona. The resulting angular grid size within this region is $0.35^{\circ}$.

Each of our simulations performs 90000 iterations with the local time stepping. The first 80000 iterations of the simulation are based on the second-order shock-capturing Linde scheme with the third-order Koren limiter \citep[][]{Koren:1993}. To achieve a higher numerical accuracy, the last 10000 iterations are performed using a fifth-order numerical scheme \citep[][]{Chen:2016} with the MP5 limiter \citep[][]{Suresh:1997}.

\section{Magnetic Field Results}\label{section_B_results}

\subsection{White-Light Images and Local Magnetic Field Structures}\label{subsection_WL_B}

After the simulations are finished, \textbf{we synthesized the Thomson-scattered white-light image for each model by integrating the scattered signal along each LOS \citep[][]{Billings:1966,Manchester2004} from the direction of Earth.} Figure~\ref{fig2_wl} shows the synthesized white-light images of the four models in panels (a)–(d), along with an eclipse image from Petr Hor\'alek\footnote{\url{https://www.petrhoralek.com/?p=24033}} in panel (e). Our synthetic images are processed with an unsharp mask filter and an exponential radial filter. \textbf{The unsharp mask filter is performed using the IDL function ``unsharp\_mask'' with an amount of $3000$.} We note that the observed image is obtained from a composite of many photographs and is presented here for reference only. The synthetic images of the four models show significant discrepancies. The following analysis will discuss how the differences in the magnetic maps lead to discrepancies in the features of the synthetic images.

\begin{figure}[tp!]
\centering {\includegraphics[width=1.0\hsize]{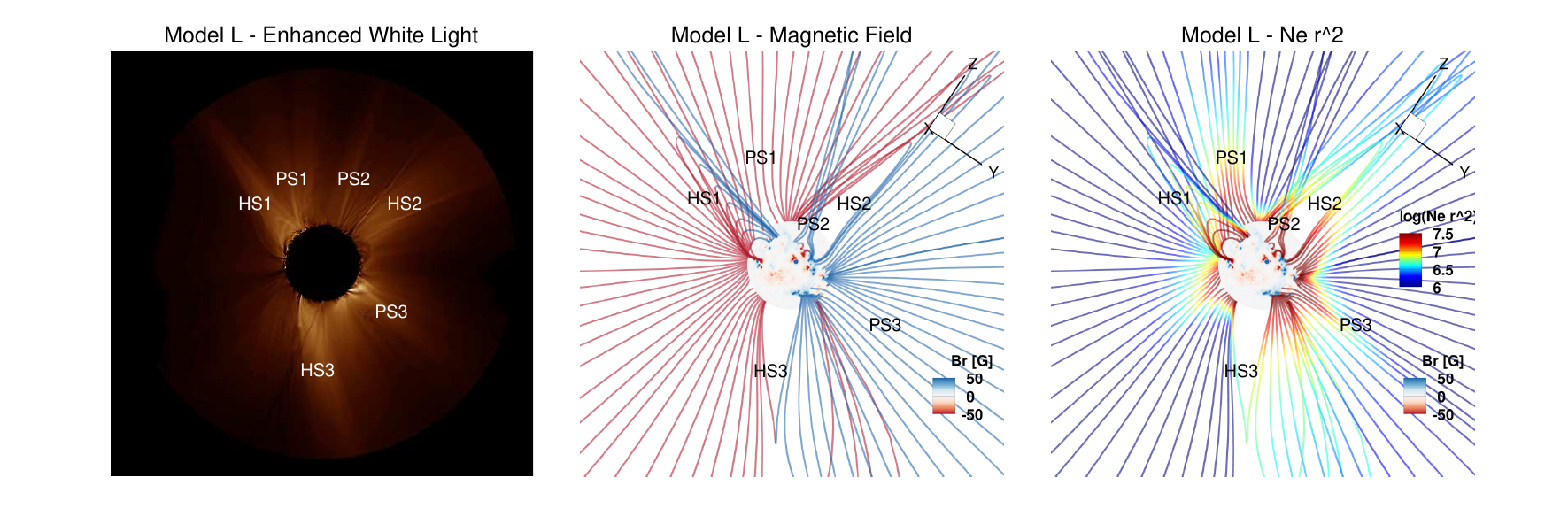}} 
\caption{The synthetic white-light image of Model L and the 3D magnetic field. Left panel: the coronagraph image synthesized from Model L. Middle panel: magnetic field lines colored according to the sign of $B_{r}$ (blue for positive and red for negative). Right panel: magnetic field lines colored according to the electron density multiplied by $r^{2}$ (to enhance contrast) on a logarithmic scale.}  \label{fig_wlcompare}
\end{figure}

Our first step is to find the magnetic field structures corresponding to the features of interest in the synthesized images. We examined the magnetic field and electron density distribution in the 3D space of each model. For each specific feature of interest in the synthetic image, we identified the magnetic structure with both a geometry consistent with this feature and a high electron density distribution. Here, we use Model L as an example to present the result. Figure~\ref{fig_wlcompare} shows the synthesized white-light image and the magnetic field lines in Model L. 
The right panel of Figure~\ref{fig_wlcompare} shows the magnetic field lines colored by $\log{(N_{e}r^{2})}$, where $N_{e}$ is the electron density. The factor $r^{2}$ is applied to enhance the contrast. We focus on six outstanding features in the synthesized white-light image, and found that they are associated with three helmet streamers (HS1, HS2, and HS3) and three pseudostreamers (PS1, PS2, and PS3). These structures are labeled in each panel. In the following subsection, we will focus on the white-light features and corresponding magnetic field structures in the upper-left and bottom parts of the synthetic images.

\begin{figure}[tp!]
\centering {\includegraphics[width=0.95\hsize]{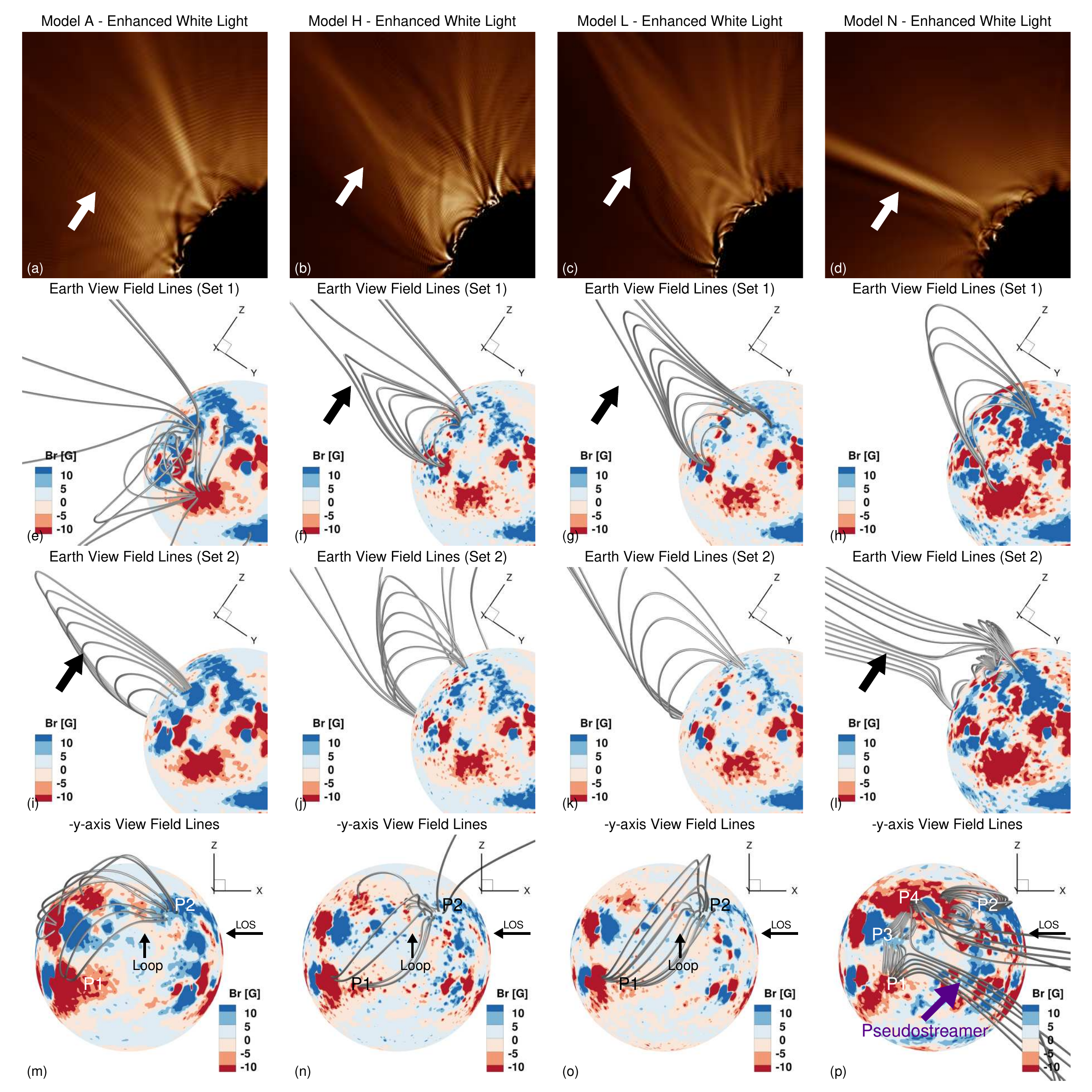}} 
\caption{Panels (a)-(d): Upper-left part of the synthesized white-light images of each model. Panels (e)-(h): The first set of magnetic field lines viewed from the Earth. Panels (i)-($\ell$): The second set of magnetic field lines viewed from the Earth's perspective. Panels (m)-(p): The second set of magnetic field lines viewed from the negative $y$-axis. The four columns represent Model A, Model H, Model L, and Model N, respectively.} \label{fig4_upleft}
\end{figure}

\subsection{Upper-left part of synthetic images}

Figure~\ref{fig4_upleft} presents the upper-left part of the white-light images and the magnetic field structures in the models. In Figure~\ref{fig4_upleft} (a)-(d), we use white arrows to highlight features of interest. These features appear as radially elongated structures. Note that the feature of interest of Model A (Figure~\ref{fig4_upleft} (a)) is relatively less prominent compared to those of the other models. The second and third rows of Figure~\ref{fig4_upleft} show two sets of magnetic field lines in each model. 
The first set (Set 1) is anchored to the magnetic poles on the disk, whereas the second set (Set 2) is anchored to the magnetic poles near the left limb. The black arrows in the second and third rows of Figure~\ref{fig4_upleft} highlight the magnetic field structures corresponding to the features highlighted with white arrows in Figure~\ref{fig4_upleft}(a)-(d). 

In both Models H and L, the feature of interest is produced by a helmet streamer within Set 1. Figure~\ref{fig4_upleft}(e) shows that Model A contains a comparatively complicated magnetic field system in Set 1. \textbf{The coronal loops in Set 1 of Model N exhibit a similar connectivity to the helmet streamers in Models H and L, while these loops do not correspond to any visible feature in Figure~\ref{fig4_upleft}(d)}. The features of interest in Figure~\ref{fig4_upleft}(a) and (d) correspond to the magnetic field structures in Set 2 of Models A and N. Figure~\ref{fig4_upleft}(i) indicates that a helmet streamer appears in Set 2 of Model A. Figure~\ref{fig4_upleft}(j) and (k) show that Models H and L contain closed loops within Set 2. The last row of Figure~\ref{fig4_upleft} shows Set 2 of each model viewed from the $-y$ axis. 
Comparing Figure~\ref{fig4_upleft}(m)-(o), we find that in Models A, H, and L the magnetic field lines within Set 2 are anchored to the same two magnetic poles (labeled by $P_{1,\,2}$). In contrast, Model N yields a much more distinct field structure, a pseudostreamer in Set 2. In Figure~\ref{fig4_upleft}(p), we label two more magnetic poles $P_{3,\,4}$. In the previous three models, magnetic poles similar to $P_{3,\,4}$ are identifiable, yet they do not play a key role in shaping the 3D field structure, while in Model N, the magnetic poles $P_{3,\,4}$ together with the pole $P_{1}$ are the foot points of the dominating magnetic field structure (i.e., the pseudostreamer). 
The magnetic field lines connecting $P_{1}$ and $P_{2}$ do not appear in Model N.

The analysis above reveals that the 3D magnetic field connectivity of Model N within Set 2 is significantly different from those of the other models. By comparing the magnetic poles $P_{1,\,2,\,3,\,4}$ of Model N with the poles in other models, we find that the most essential distinction is that $P_{4}$ contains significantly more magnetic flux than its counterparts in the other models. This comparison is consistent with the interpretation from Figure~\ref{fig_compare_maps} that the NSO-NRT-HMI map has a stronger average field strength in high-latitude regions. The key effect of a strong pole $P_{4}$ is that its intense negative flux breaks the direct connectivity between $P_{1,\,2}$. The two field structures associated with $P_{4}$, i.e, the pseudostreamer with foot points $P_{1,\,3,\,4}$ and the loops connecting $P_{4}$ and $P_{2}$, can be interpreted as the resulting structure due to this break of connectivity between $P_{1,\,2}$. Based on the analysis above, we point out that despite the similar identifiable magnetic poles in all the magnetic maps, the enhancement of a single magnetic pole can significantly modify the 3D magnetic field connectivity and produce distinct structures. 

\begin{figure}[tp!]
\centering {\includegraphics[width=0.95\hsize]{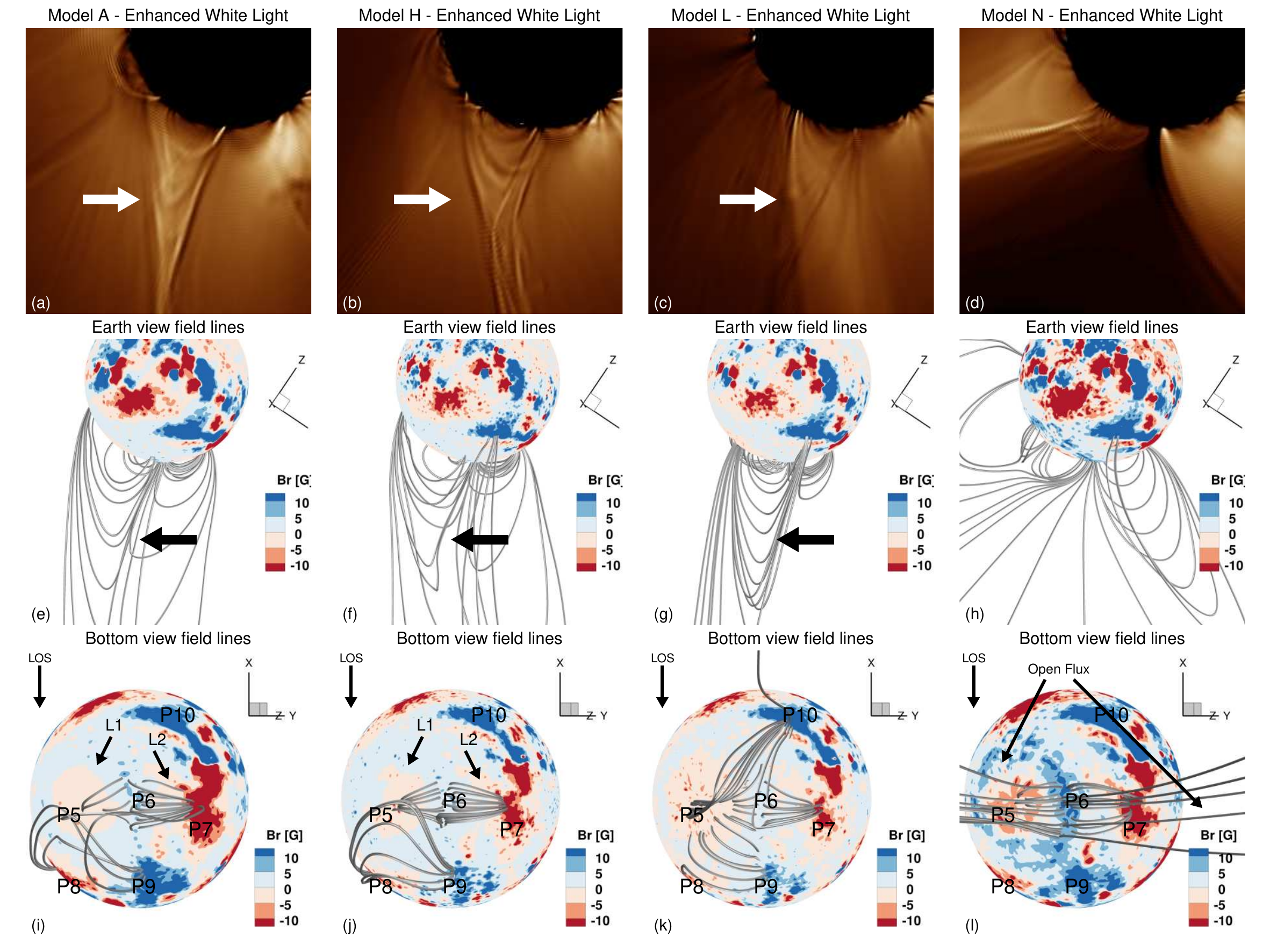}} 
\caption{Panels (a)-(d): Bottom part of the synthesized white-light images of each model. Panels (e)-(h): The magnetic field lines viewed from the Earth. Panels (i)-($\ell$): The magnetic field lines viewed from a bottom perspective.} \label{fig5_bottom}
\end{figure}

\subsection{Bottom part of synthetic images}

Figure~\ref{fig5_bottom} shows the bottom part of the white-light images and corresponding magnetic structures. Each of Models A, H, and L exhibits a narrow and near-vertical feature in the white-light images, although the geometry and brightness vary across models. These structures are associated with the helmet streamers, as shown in Figure~\ref{fig5_bottom}(e)-(g). An intuitive expectation is that the foot points of the helmet streamers are homologous magnetic poles in the three models. However, the bottom view of the field lines (Figure~\ref{fig5_bottom}(i)-(k)) contradicts this expectation. In the last row of Figure~\ref{fig5_bottom}, we label six magnetic poles ($P_{5-10}$) that are identifiable in each model. The loops $L_{1}$ and $L_{2}$ that connect the positive pole $P_{6}$ and negative poles $P_{5,7}$ are visible in all models. \textbf{We notice that} the foot points of the helmet streamer are $P_{8}$ and $P_{9}$ in Model A and Model H, while in Model L the helmet streamer is anchored to $P_{5}$ and $P_{10}$. The field lines connecting $P_{8}$ and $P_{9}$ in Model L have a low height and are not associated with the helmet streamer. 

Compared with other models, Model N again exhibits distinct features. Figure~\ref{fig5_bottom}(d) shows a large region with relatively weak white-light intensity. Figure~\ref{fig5_bottom}(h) and $(\ell)$ indicate that the weak signal is due to the expanding open magnetic flux originating from $P_{6}$. The $P_{6}$ in Model N contains remarkably more positive magnetic flux than the $P_{6}$ poles of other models, thus producing the open flux region that shields the surrounding structures. When comparing the magnetic poles of Model N with all other maps, the most outstanding difference is the stronger poles $P_{5,6}$ in Model N, which directly leads to the open flux region discussed above. We note that this is again consistent with Figure~\ref{fig_compare_maps}.

\begin{figure}[tp!]
\centering {\includegraphics[width=1.0\hsize]{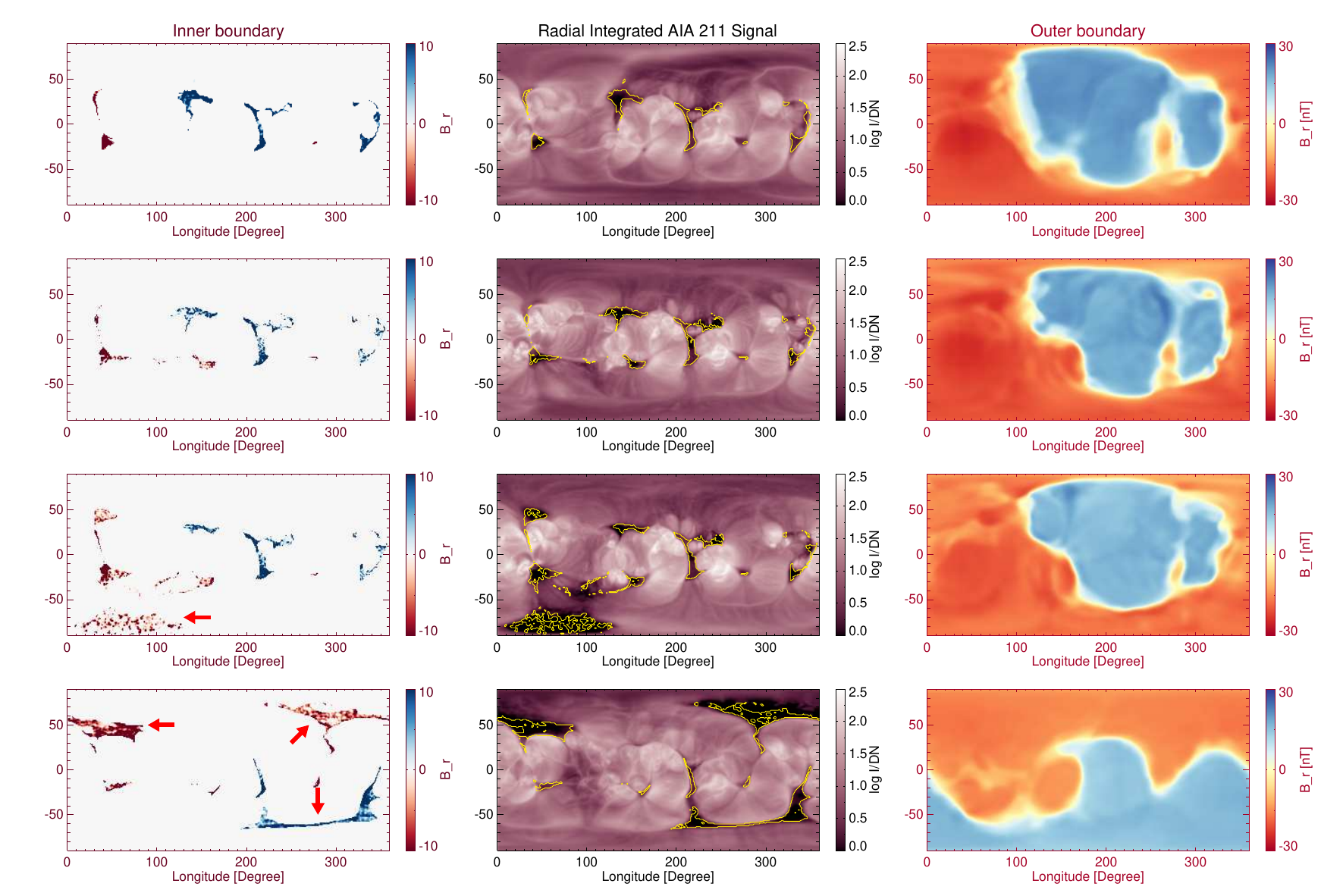}} 
\caption{\textbf{First panel: Open flux regions at the inner boundary, colored by $B_{r}$ (blue/red).} Second panel: The full-sun maps of the synthetic AIA 211 {\AA} emission. Last panel: The $B_{r}$ distribution at $r=24\; R_s$. The four rows correspond to the Model A, Model H, Model L, and Model N, respectively} \label{fig_ch}
\end{figure}

\begin{table*}[ht!]
\caption{Areas ($A$) and fluxes ($\Phi$) of open flux regions (O) and unique open flux regions (U) of each model.} \label{tab_ch}
\begin{center}
\setlength{\tabcolsep}{0.25cm}
\begin{tabular}{ccccccc}
\hline\hline
Model & $A_{O}$ [$10^{21}$ cm$^{2}$] & $A_{U}$ [$10^{21}$ cm$^{2}$] & $A_{U}/A_{O}$ & $\Phi_{O}$ [$10^{21}$ Mx] & $\Phi_{U}$ [$10^{21}$ Mx] & $\Phi_{U}/\Phi_{O}$ \\ \hline
Model A & $1.66$ & $-$ & $-$ & $62.9$ & $-$ & $-$ \\
Model H & $1.97$ & $-$ & $-$ & $61.3$ & $-$ & $-$ \\
Model L & $2.67$ & $0.334$ & $12\%$ & $51.7$ & $1.97$ & $3.8\%$ \\
Model N & $3.22$ & $1.34$ & $42\%$ & $47.8$ & $17.2$ & $36\%$ \\
\hline
\end{tabular}
\end{center}
\end{table*}

\subsection{Open Magnetic Flux}\label{subsection_open_B}

Another aspect of the magnetic field we are interested in is the open magnetic flux. To locate the open flux regions of each model, we trace along the magnetic field lines from an angular-uniform grid at the inner boundary \textbf{with an angular resolution of $1^{\circ}$}, and determine whether each magnetic field connects to the outer boundary (open) or returns to the inner boundary (closed). The first column in Figure~\ref{fig_ch} presents the results of field line tracing. \textbf{The open flux regions are colored by $B_{r}$, while the closed flux regions remain white.} The open flux regions do not include the north and south poles ($\lambda=\pm 90^{\circ}$) in any of the four models, which is consistent with the weak solar polar dipole field near the solar maximum. \textbf{In the second column}, we show the full-sun maps of the radially integrated AIA 211 {\AA} emission for reference. 
The synthesis of the AIA 211 {\AA} emission used the same CHIANTI version and \textbf{elemental} abundance as were used for the radiative cooling in the simulation. The yellow contours indicate the boundaries of the open flux regions that match well the regions with low AIA 211 {\AA} emissions (i.e., coronal holes) in agreement with earlier studies \citep[e.g.,][]{Krieger1973,Bohlin1977,Cranmer2009}. 

For each model, we use red arrows to label those {\em unique} open flux regions that do not have a similar counterpart in any other model. To quantify the properties of the open flux regions, we use the $A_{O}$ ($\Phi_{O}$) to denote their total area (unsigned flux), and $A_{U}$ ($\Phi_{U}$) to denote the total area (unsigned flux) of the unique portions. 
These quantities are listed in Table~\ref{tab_ch} for the four simulations. We note that the absence of unique open flux regions in Models A and H does not mean that they contain identical open flux regions. Generally, Models A, H, and L exhibit similar open flux regions in the low-latitude regions ($|\lambda|<40^{\circ}$), while the two major differences are that (1) Model A does not capture the open flux region near $\phi=160^\circ, \lambda=-30^\circ$ (which appears in both Model H and L), and (2) Model L exhibits a unique open flux region near the southern pole. Table~\ref{tab_ch} suggests that \textbf{the unique open flux regions in Model L contain only a small portion of the area and the unsigned flux}. The open flux regions in Model N exhibit significant differences from those in the other three models. First, Model N contains almost no open flux regions within $100^{\circ}<\phi<200^{\circ}$. Second, Model N contains two large-scale unique open flux regions, one near $\lambda=50^{\circ}$ (which extends across the $0^{\circ}/360^{\circ}$ longitudinal boundary) and the other near $\lambda=-50^{\circ}$. These two unique regions contain $36\%$ of the open magnetic flux in Model N. 
We note that the unique open flux regions only capture a part of the differences among the open flux regions of the models. 

The last column of Figure~\ref{fig_ch} shows the $B_{r}$ distribution at the outer boundary (24\,$R_s$) of each model. 
Models A, H, and L exhibit similar patterns: the open flux is dominated by an equatorial dipole field. 
In contrast, Model N exhibits a heliospheric current sheet that extends continuously around the Sun in the longitudinal direction. This pattern indicates that the open magnetic flux is dominated by an axial dipole in Model N.
The hemispheric polarity of this axial dipole is consistent with the hemispheric polarity of the unique open flux regions of Model N. It is interesting to note that this hemispheric polarity is opposite to the hemispheric polarity of the field near $|\lambda|=\pm 90^{\circ}$ at the inner boundary of Model N (negative in the south pole, see bottom right panel of Figure~\ref{fig1_map}). When the strong axial dipole dominates the solar magnetic field at the solar surface (e.g., during solar minimum), the open flux regions primarily locate near the northern and southern poles. Thus, the magnetic polarity in regions near $|\lambda|=\pm 90^{\circ}$ determines the polarity of the open flux. However, this relation does not apply to the cases with a complicated global field (e.g., near the solar maximum). 
From the results above, we learn that the open flux polarity is sensitive to the global photospheric flux distribution near solar maximum, rather than the polarity near the north and south poles.

\section{Coronal Heating}\label{section_heating}

\begin{figure}[tp!]
\centering {\includegraphics[width=0.5\hsize]{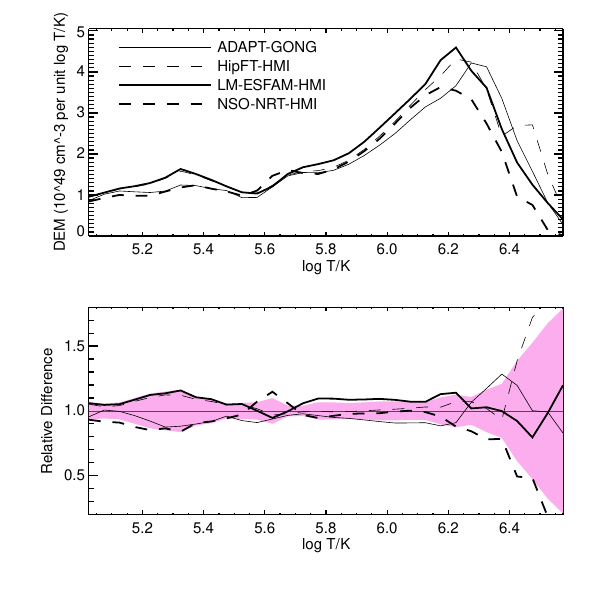}} 
\caption{Upper panel: The DEM as a function of $\log T_{e}$ in each model. Lower panel: The relative difference of each DEM curve to the average. The red horizontal line indicates the average DEM distribution, of which the relative difference is constantly zero. The pink region indicates the standard deviation range for each $\log{T_{e}}$ value.} \label{fig_EM_compare}
\end{figure}

In this section, we assess the impact of using different maps on the properties of coronal heating. We calculate the volume-integrated total heat dissipation rate $P_{D}=\sum_{i}Q_{\mathrm{heat},i}V_{i}$. Here $Q_{\mathrm{heat},i}$ is the Aflv\'en wave energy dissipation rate per unit volume and $V_{i}$ is the volume of the $i$th simulation cell. The values of $P_{D}$, along with the ratio between the total dissipation rate and the total injection rate ($P_{D}/P_{i}$) of each model, are shown in Table\ref{tab_unsigned_flux}. The result indicates that, in general, only approximately half of the injected Alfv\'en wave energy is dissipated in the domain. The rest of the injected Alfv\'en wave energy is mainly lost by Alfv\'en waves propagating through the inner and outer boundaries. The maximum value ($1.73\times10^{28}$ erg s$^{-1}$) and the minimum value ($1.47\times10^{28}$ erg s$^{-1}$) of $P_{D}$ yield a relative difference of $\approx 18\%$, which is considerably smaller than the original relative difference in the total unsigned flux in different maps ($86\%$). \textbf{We note that this is expected, as we used different $(S_\mathrm{A}/B)_{\odot}$ values for the four maps to compensate for the difference in the total unsigned flux and thereby ensure that the energy injection rate $P_{I}$ is identical across all simulations.}

It is interesting to examine the variation of the differential emission measures (DEMs) in the models as a result of coronal heating. \textbf{For our purposes}, we define the DEM in the 3D space as the distribution of the \textbf{volume} emission measure (EM) as a function of the logarithm of the electron temperature ($\log{T_{e}}$ [K]):
\begin{equation}
   \mathrm{DEM}(\log{T_{e}})=\frac{\Delta \mathrm{EM}}{\Delta \log{T_{e}}}
\end{equation}
where $\mathrm{EM}$ in volume $V$ is given by $\mathrm{EM}=\int_{V}N_{e}^{2}dV$. DEM is useful in describing the emitting plasma in the optically thin corona. 
Figure~\ref{fig_EM_compare} presents $\mathrm{DEM}(\log{T_{e}})$ for each model. We calculate the average $\overline{\mathrm{DEM}}(\log{T_{e}})$ value of the four models and derive the relative difference $(\mathrm{DEM}-\overline{\mathrm{DEM}})/\overline{\mathrm{DEM}}$ for each model, 
which is shown in the second panel of Figure~\ref{fig_EM_compare}. We also plot the standard deviation of the relative differences as a pink region between $\pm \sigma(\log{T_{e}})$.
The relative difference across the four $\mathrm{DEM}$ curves is generally smaller than $15\%$ for $5.8<\log{T_e}<6.2$, 
which is the typical temperature range in the quiet sun regions. However, a significant discrepancy among the $\mathrm{DEM}$ curves can be found for $\log{T_{e}}>6.4$. These temperature values mainly occur near the active regions. An explanation of this large discrepancy would require a closer look at local coronal loops and the Alfv\'en wave propagation and dissipation \citep[e.g.,][]{Shi2024}. We refrain from doing this analysis here due to its complexity.

\section{Discussions}\label{section_discussion}

We used four photospheric magnetic maps to perform MHD simulations of the corona during the 8 April 2024 Total Solar Eclipse. We remapped the four original $B_{r}$ maps onto the same $1^\circ\times1^\circ$ grid. We compared these maps and found a relative difference of $86\%$ in the total unsigned flux. We used different $(S_\mathrm{A}/B)_{\odot}$ values for the four derived maps to keep the total Alfv\'en wave energy injection rate $P_{I}$ the same in all the models. The resulting total Alfv\'en wave energy dissipation rate $P_{D}$ varies only about $18\%$. We synthesized the white-light images from the models and found the magnetic field structures corresponding to the features of interest in the synthetic images. The synthetic white-light images reveal significant discrepancies across the four models, which are due to differences in the $B_{r}$ maps. We chose two areas in these models and analyzed the connection between the simulation results and the maps. We then compared the open flux regions of the four models. We also analyzed the unique open flux regions in each model and evaluated the area and magnetic flux in those regions. We also analyzed the Alfv\'en wave energy dissipation rate in each model and compared the DEM distributions across the models.

One of our key findings is that the 3D magnetic field topology is highly sensitive to the local distribution of the magnetic flux in the $B_{r}$ map: even when the maps exhibit similar polarity distribution patterns, differences in the magnetic flux distributions can still produce distinct magnetic field topologies. We also find that the polarity of the open magnetic flux depends on the choice of magnetic maps. Another key finding of this work is that the coronal heating in the high-temperature regions is especially sensitive to the use of different maps. The $\mathrm{DEM}$ distribution exhibits relatively small variation ($\approx 15\%$) across the models for the typical coronal temperature, while exhibiting significant variations ($>40\%$) for $T_{e}>2.5\,$MK. Further analysis of local Alfv\'en wave propagation and dissipation is needed to provide a clear explanation of this dependence. 

We find significant differences among the models in our comparisons of the simulation results. Among the four models, Model N exhibits the most distinctive results. The comparison of the photospheric $B_{r}$ maps in Figure~\ref{fig_compare_maps} suggests that the NSO-NRT-HMI map has a significantly stronger field strength in the high-latitude regions than the other maps, which can presumably explain most of the unique features in the results of Model N. Our analysis in Subsection~\ref{subsection_WL_B} shows that the stronger field strength in the high-latitude regions in Model N leads to a unique local balance of the magnetic polarity, which results in magnetic structures that are different from those in other models. The stronger high-latitude field strength also accounts for the unique open flux regions of Model N in Figure~\ref{fig_ch}. Despite the distinctive simulation results of Model N, the other three models also exhibit significant noticeable differences, especially in the local magnetic field structures. 

On the other hand, Models H and L show several similarities in the magnetic field structures. For example, our analysis of Figure~\ref{fig4_upleft} suggests that both models contain a helmet streamer in the upper left part of the field-of-view. Figure~\ref{fig_ch} shows that the two models contain comparable open flux regions and exhibit similar $B_{r}$ distributions at the outer boundary. The similarities of the two models are consistent with Figure~\ref{fig_compare_maps}, in which the difference of $\bar{B_{r}}(\lambda_{j})$ between the two models is $<1$\, G for a wide range of latitudes.

\textbf{However, our work does not address how the underlying methods used to produce the full-Sun magnetic maps account for the resulting differences in the maps.} There are several differences in the production of the four maps: (1) The ADAPT-GONG map is based on the GONG measurements of the magnetic field, while the other maps are based on the SDO/HMI measurements. \textbf{(2) The ADAPT-GONG, HipFT-HMI and LM-ESFAM-HMI maps include the flux transport models, while the NSO-NRT-HMI does not. (3) The flux transport models in the ADAPT-GONG and HipFT-HMI maps are based on the Eulerian scheme, while the flux transport model in the LM-ESFAM-HMI map is based on the Lagrangian scheme.} (4) The NSO-NRT-HMI map is based on the vector field \textbf{magnetograms}, while the other maps are derived from the LOS component of the magnetic field. (5) Although both the HipFT-HMI and LM-ESFAM-HMI maps are based on the HMI LOS magnetogram, they use different processing methods. \textbf{Future work is needed to connect the differences in the production methods with the differences in the resulting maps.} For example, it is yet to be answered whether the stronger field strength in the high-latitudes regions of the NSO-NRT-HMI map is due to the use of the vector field \textbf{magnetograms} or due to other factors.

Our findings underscore the critical role of accurate photospheric magnetic maps in coronal modeling. The discrepancies among maps are due to different photospheric measurement instruments and data processing methods applied to these maps. We note that the diversity in processing methods is, to some extent, a consequence of the absence of simultaneous measurements of the magnetic field over the entire photosphere. If we had instruments to observe the entire solar surface (and assuming they are well calibrated), there would be much less need to assimilate the data over a range of time. \citet{Downs2025} used the observation from both the SDO/HMI and the Polarimetric and Helioseismic Imager \citep[PHI,][]{Solo_PHI} on board the Solar Orbiter mission \citep[SolO,][]{Solo}, which provides the photospheric magnetic field measurement from another viewing angle. Future instruments will also help address this problem. For example, the Vigil mission \citep[][]{VigilESA}, led by the European Space Agency (ESA), will provide observations from the Sun-Earth L5 Lagrange point. Recently, the use of deep learning to derive photospheric magnetograms on the far side has been viewed as an alternative approach \citep[e.g.,][]{Kim2019, Jeong2022}. Despite the dramatic growth of this method, more studies are still required to test its reliability or apply it to coronal models. From a long-term perspective, an optimal approach would be future missions capable of providing continuous, simultaneous, full-disk and direct observations from multiple viewing angles, especially the far-side and polar regions. 

\section*{Acknowledgments}
This work is primarily supported by the NASA LWS Strategic Capabilities grant 80NSSC22K0892 (SCEPTER) and the NASA SWxC grant 80NSSC23M0191 (CLEAR). 
We thank the \textit{SDO}/HMI science project for providing the original full-disk magnetograms used to create synoptic maps in this study. The authors thank the HipFT model development team at Predictive Science Inc. We thank the ADAPT model development team, supported by the Air Force Research Laboratory (AFRL) and the Air Force Office of Scientific Research (AFOSR) tasks 18RVCOR126 and 22RVCOR012. Data used in the ADAPT model were acquired by GONG instruments operated by NISP/NSO/AURA/NSF with a contribution from NOAA. The NSO is operated by the Association of Universities for Research in Astronomy (AURA), Inc. under a cooperative agreement with the National Science Foundation. In this paper, the views expressed are those of the authors and do not reflect the official guidance or position of the Department of Defense (DoD), the United States Air Force, or the National Aeronautics and Space Administration. We thank Dr. Spiro Antiochos for valuable discussions and insightful suggestions on this work. We also acknowledge the high-performance computing support from the Texas Advanced Computing Center (TACC) Frontera at the University of Texas at Austin \citep{stanzione2020frontera}, and the NASA supercomputing system Pleiades as part of the NASA High-End Computing (HEC) Program through the NASA Advanced Supercomputing (NAS) Division at Ames Research Center.

\setcounter{figure}{0}
\renewcommand{\thefigure}{A\arabic{figure}}
\setcounter{table}{0}
\renewcommand{\thetable}{A\arabic{table}}
\setcounter{equation}{0}
\renewcommand{\theequation}{A\arabic{equation}}
\appendix

\section{The HipFT Model} \label{appendix_hipft}

The HipFT model, developed by Predictive Science Inc., involves the data assimilation of photospheric magnetic field observations and the simulation of the photospheric $B_{r}$ evolution \citep[][]{Caplan2025,Downs2025}. In this work, we used the version of HipFT from the GitHub repository as of February 2024.

We used the HMI NRT $B_{r}$ magnetograms as the input for HipFT. Each original full-disk magnetogram was remapped onto a uniform Carrington grid ($1024\times512$ in longitude and latitude), which was later used as the grid in the HipFT simulation. During the HipFT simulation, the program performs data assimilation at a cadence of 1 hour of simulation time. Each assimilation process merges the input magnetogram into the simulated $B_{r}$ map with an assimilation weight distribution $F$:
\begin{equation}
    F=\begin{cases}
        \mu^{4},& \theta_{d}<80^{\circ} \\
        0,& \theta_{d}\geq80^{\circ}
    \end{cases},
\end{equation}
\begin{equation}
    \mu=\cos{\theta_{d}}.
\end{equation}
Here $\theta_{d}$ is the heliocentric angle. We note that the HipFT repository has undergone several updates since February 2024; therefore, the version used here does not include the flux-balance feature described by \citet{Caplan2025}, which is included in the latest version of HipFT.

The HipFT simulation includes the flux transport and diffusion processes. The flux transport process encompasses solar differential rotation, meridional flow, and supergranular flow. The simulation of the supergranular flow incorporates a series of random convective velocity maps produced by the Convective Flow Generator (ConFlow\footnote{\url{https://github.com/predsci/conflow}}). In this work, we did not perform the ConFlow calculations; instead, we adopted the existing sequence of the convective velocity maps provided by the HipFT sample input dataset\footnote{\url{https://zenodo.org/records/11205509}}. For the flux diffusion process, we used the default value of the diffusion coefficient constant. We note that our version of HipFT does not include the random flux emergence process, which is implemented in the latest version of HipFT \citep[][]{Caplan2025}.

We performed data assimilation and simulation using magnetograms from 2024 March 5 to 2024 April 7. The HMI NRT magnetogram at 23:58:33 UT on 2024 March 4 is used as the input, and the simulated $B_{r}$ map at 00:00:00 UT on 2024 April 7 was retrieved as the output.

\section{The LM-ESFAM Model} \label{appendix_lmsal}

The ESFAM model, as originally developed at Lockheed Martin, is described in \citet{2001ApJ...547..475S}. This ``pure simulation" model enabled the evolution of the photospheric magnetic field of cool Sun-like stars to be modeled. When run with the parameter set applicable for the Sun, the model provides a time series of full-Sun magnetic maps throughout multiple sunspot cycles. The emergence of flux onto the model photosphere is based on empirical descriptions that apply to the full spectrum of bipoles, ranging from active regions on the large end to ephemeral regions on the small end. Once flux has been inserted into the model, the flux is evolved in time due to parametrized descriptions of the large-scale differential rotation and meridional flow patterns. The dispersal of flux from convective flow patterns is also captured. 

In assimilation mode, described in \citet{2003SoPh..212..165S}, observed magnetogram images from HMI are inserted into the model in order to capture the actual locations of bipolar regions. The state of the assimilation model is sampled every six hours, and the magnetic maps at these times are publicly available via \citet{derosa_2025_lm_fasam_hmi}. For the study presented here, starting about a month prior to the date of the TSE, the assimilation model was run forward in time until the TSE date, at which time the state of the model was captured.


\bibliography{ref,ref_SWIFT_2}{}

@ARTICLE{Alfven1942,
       author = {{Alfv{\'e}n}, H.},
        title = "{Existence of Electromagnetic-Hydrodynamic Waves}",
      journal = {\nat},
         year = 1942,
        month = oct,
       volume = {150},
       number = {3805},
        pages = {405-406},
          doi = {10.1038/150405d0},
       adsurl = {https://ui.adsabs.harvard.edu/abs/1942Natur.150..405A}
}

@ARTICLE{Parker1983a,
       author = {{Parker}, E.~N.},
        title = "{Magnetic neutral sheets in evolving fields. I - General theory.}",
      journal = {\apj},
         year = 1983,
        month = jan,
       volume = {264},
        pages = {635-647},
          doi = {10.1086/160636},
       adsurl = {https://ui.adsabs.harvard.edu/abs/1983ApJ...264..635P}
}

@ARTICLE{Parker1983b,
       author = {{Parker}, E.~N.},
        title = "{Magnetic Neutral Sheets in Evolving Fields - Part Two - Formation of the Solar Corona}",
      journal = {\apj},
         year = 1983,
        month = jan,
       volume = {264},
        pages = {642},
          doi = {10.1086/160637},
       adsurl = {https://ui.adsabs.harvard.edu/abs/1983ApJ...264..642P}
}

@ARTICLE{Mikic2018,
       author = {{Miki{\'c}}, Zoran and {Downs}, Cooper and {Linker}, Jon A. and {Caplan}, Ronald M. and {Mackay}, Duncan H. and {Upton}, Lisa A. and {Riley}, Pete and {Lionello}, Roberto and {T{\"o}r{\"o}k}, Tibor and {Titov}, Viacheslav S. and {Wijaya}, Janvier and {Druckm{\"u}ller}, Miloslav and {Pasachoff}, Jay M. and {Carlos}, Wendy},
        title = "{Predicting the corona for the 21 August 2017 total solar eclipse}",
      journal = {Nature Astronomy},
         year = 2018,
        month = aug,
       volume = {2},
        pages = {913-921},
          doi = {10.1038/s41550-018-0562-5},
       adsurl = {https://ui.adsabs.harvard.edu/abs/2018NatAs...2..913M}
}

@ARTICLE{Habbal:2007,
       author = {{Habbal}, Shadia Rifai and {Morgan}, Huw and {Johnson}, Judd and {Arndt}, Martina Belz and {Daw}, Adrian and {Jaeggli}, Sarah and {Kuhn}, Jeff and {Mickey}, Don},
        title = "{Localized Enhancements of Fe$^{+10}$ Density in the Corona as Observed in Fe XI 789.2 nm during the 2006 March 29 Total Solar Eclipse}",
      journal = {\apj},
         year = 2007,
        month = jul,
       volume = {663},
       number = {1},
        pages = {598-609},
          doi = {10.1086/518403}}

@ARTICLE{Habbal:2013,
       author = {{Habbal}, S. Rifai and {Morgan}, H. and {Druckm{\"u}ller}, M. and {Ding}, A. and {Cooper}, J.~F. and {Daw}, A. and {Sittler}, E.~C.},
        title = "{Probing the Fundamental Physics of the Solar Corona with Lunar Solar Occultation Observations}",
      journal = {\solphys},
         year = 2013,
        month = jul,
       volume = {285},
       number = {1-2},
        pages = {9-24},
          doi = {10.1007/s11207-012-0115-5}}

@ARTICLE{Lionello2009,
       author = {{Lionello}, Roberto and {Linker}, Jon A. and {Miki{\'c}}, Zoran},
        title = "{Multispectral Emission of the Sun During the First Whole Sun Month: Magnetohydrodynamic Simulations}",
      journal = {\apj},
         year = 2009,
        month = jan,
       volume = {690},
       number = {1},
        pages = {902-912},
          doi = {10.1088/0004-637X/690/1/902},
       adsurl = {https://ui.adsabs.harvard.edu/abs/2009ApJ...690..902L}
}

@Article{vanderholst:2014,
  author     = {B. {van der Holst} and I. V. Sokolov and X. Meng and M. Jin and W. B. Manchester and G. Toth and T. I. Gombosi},
  title      = {Alfv{\'e}n Wave Solar Model {(AWSoM)}: {C}oronal Heating},
  journal    = {\apj},
  year       = {2014},
  volume     = {782},
  pages      = {81},
  doi        = {10.1088/0004-637X/782/2/81},
}

@ARTICLE{Jin2012,
       author = {{Jin}, M. and {Manchester}, W.~B. and {van der Holst}, B. and {Gruesbeck}, J.~R. and {Frazin}, R.~A. and {Landi}, E. and {Vasquez}, A.~M. and {Lamy}, P.~L. and {Llebaria}, A. and {Fedorov}, A. and {Toth}, G. and {Gombosi}, T.~I.},
        title = "{A Global Two-temperature Corona and Inner Heliosphere Model: A Comprehensive Validation Study}",
      journal = {\apj},
         year = 2012,
        month = jan,
       volume = {745},
       number = {1},
          eid = {6},
        pages = {6},
          doi = {10.1088/0004-637X/745/1/6},
       adsurl = {https://ui.adsabs.harvard.edu/abs/2012ApJ...745....6J}
}

@article{henney2012forecasting,
  title={{Forecasting F10.7 with Solar Magnetic Flux Transport Modeling}},
  author={Henney, CJ and Toussaint, WA and White, SM and Arge, CN},
  journal={Space Weather},
  volume={10},
  number={2},
  year={2012},
  doi={10.1029/2011SW000748},
  publisher={Wiley Online Library}
}

@article{arge2010air,
  title={{Air Force Data Assimilative Photospheric Flux Transport (ADAPT) Model}},
  author={Arge, C Nick and Henney, Carl J and Koller, Josef and Compeau, C Rich and Young, Shawn and MacKenzie, David and Fay, Alex and Harvey, John W},
  journal={AIP conference proceedings},
  volume={1216},
  number={1},
  pages={343--346},
  year={2010},
  doi={10.1063/1.3395870},
  organization={American Institute of Physics}
}

@article{arge2011improving,
  title={{Improving Data Drivers for Coronal and Solar Wind Models}},
  author={Arge, C Nick and Henney, Carl J and Koller, Josef and Toussaint, W Alex and Harvey, John W and Young, Shawn},
  journal={5th international conference of numerical modeling of space plasma flows (astronum 2010)},
  volume={444},
  pages={99},
  adsurl = {https://ui.adsabs.harvard.edu/abs/2011ASPC..444...99A},
  year={2011}
}

@inproceedings{harvey1998new,
  title={New types and uses of synoptic maps},
  author={Harvey, J and Worden, J},
  booktitle={Synoptic Solar Physics},
  volume={140},
  pages={155},
  year={1998}
}

@article{bertello2014uncertainties,
  title={Uncertainties in solar synoptic magnetic flux maps},
  author={Bertello, L and Pevtsov, AA and Petrie, GJD and Keys, D},
  journal={\solphys},
  volume={289},
  pages={2419--2431},
  year={2014},
  doi={10.1007/s11207-014-0480-3},
  publisher={Springer}
}

@ARTICLE{pesnell2012solar,
       author = {{Pesnell}, W. Dean and {Thompson}, B.~J. and {Chamberlin}, P.~C.},
        title = "{The Solar Dynamics Observatory (SDO)}",
      journal = {\solphys},
         year = 2012,
        month = jan,
       volume = {275},
       number = {1-2},
        pages = {3-15},
          doi = {10.1007/s11207-011-9841-3},
       adsurl = {https://ui.adsabs.harvard.edu/abs/2012SoPh..275....3P}
}

@article{scherrer2012helioseismic,
  title={The helioseismic and magnetic imager (HMI) investigation for the solar dynamics observatory (SDO)},
  author={Scherrer, Philip Hanby and Schou, Jesper and Bush, RI and Kosovichev, AG and Bogart, RS and Hoeksema, JT and Liu, Y and Duvall, TL and Zhao, J and Title, AM and others},
  journal={\solphys},
  volume={275},
  pages={207--227},
  year={2012},
  doi={10.1007/s11207-011-9834-2},
  publisher={Springer}
}

@ARTICLE{Powell1999,
       author = {{Powell}, Kenneth G. and {Roe}, Philip L. and {Linde}, Timur J. and {Gombosi}, Tamas I. and {De Zeeuw}, Darren L.},
        title = "{A Solution-Adaptive Upwind Scheme for Ideal Magnetohydrodynamics}",
      journal = {Journal of Computational Physics},
         year = 1999,
        month = sep,
       volume = {154},
       number = {2},
        pages = {284-309},
          doi = {10.1006/jcph.1999.6299},
       adsurl = {https://ui.adsabs.harvard.edu/abs/1999JCoPh.154..284P}
}

@ARTICLE{Toth2011,
       author = {{T{\'o}th}, G{\'a}bor and {van der Holst}, Bart and {Huang}, Zhenguang},
        title = "{Obtaining Potential Field Solutions with Spherical Harmonics and Finite Differences}",
      journal = {\apj},
         year = 2011,
        month = may,
       volume = {732},
       number = {2},
          eid = {102},
        pages = {102},
          doi = {10.1088/0004-637X/732/2/102},
archivePrefix = {arXiv},
       eprint = {1104.5672},
 primaryClass = {astro-ph.SR},
       adsurl = {https://ui.adsabs.harvard.edu/abs/2011ApJ...732..102T}
}

@ARTICLE{BDP2007,
       author = {{De Pontieu}, B. and {McIntosh}, S.~W. and {Carlsson}, M. and {Hansteen}, V.~H. and {Tarbell}, T.~D. and {Schrijver}, C.~J. and {Title}, A.~M. and {Shine}, R.~A. and {Tsuneta}, S. and {Katsukawa}, Y. and {Ichimoto}, K. and {Suematsu}, Y. and {Shimizu}, T. and {Nagata}, S.},
        title = "{Chromospheric Alfv{\'e}nic Waves Strong Enough to Power the Solar Wind}",
      journal = {\sci},
         year = 2007,
        month = dec,
       volume = {318},
       number = {5856},
        pages = {1574},
          doi = {10.1126/science.1151747},
       adsurl = {https://ui.adsabs.harvard.edu/abs/2007Sci...318.1574D}
}

@ARTICLE{Tomczyk2007,
       author = {{Tomczyk}, S. and {McIntosh}, S.~W. and {Keil}, S.~L. and {Judge}, P.~G. and {Schad}, T. and {Seeley}, D.~H. and {Edmondson}, J.},
        title = "{Alfv{\'e}n Waves in the Solar Corona}",
      journal = {\sci},
         year = 2007,
        month = aug,
       volume = {317},
       number = {5842},
        pages = {1192},
          doi = {10.1126/science.1143304},
       adsurl = {https://ui.adsabs.harvard.edu/abs/2007Sci...317.1192T}
}

@ARTICLE{Sachdeva2023,
       author = {{Sachdeva}, Nishtha and {Manchester}, Ward B., IV and {Sokolov}, Igor and {Huang}, Zhenguang and {Pevtsov}, Alexander and {Bertello}, Luca and {Pevtsov}, Alexei A. and {Toth}, Gabor and {van der Holst}, Bart and {Henney}, Carl J.},
        title = "{Solar Wind Modeling with the Alfv{\'e}n Wave Solar atmosphere Model Driven by HMI-based Near-real-time Maps by the National Solar Observatory}",
      journal = {\apj},
         year = 2023,
        month = aug,
       volume = {952},
       number = {2},
          eid = {117},
        pages = {117},
          doi = {10.3847/1538-4357/acda87},
archivePrefix = {arXiv},
       eprint = {2212.05138},
 primaryClass = {astro-ph.SR},
       adsurl = {https://ui.adsabs.harvard.edu/abs/2023ApJ...952..117S}
}

@ARTICLE{Solo_PHI,
       author = {{Solanki}, S.~K. and {del Toro Iniesta}, J.~C. and {Woch}, J. and {Gandorfer}, A. and {Hirzberger}, J. and {Alvarez-Herrero}, A. and {Appourchaux}, T. and {Mart{\'\i}nez Pillet}, V. and {P{\'e}rez-Grande}, I. and {Sanchis Kilders}, E. and {Schmidt}, W. and {G{\'o}mez Cama}, J.~M. and {Michalik}, H. and {Deutsch}, W. and {Fernandez-Rico}, G. and {Grauf}, B. and {Gizon}, L. and {Heerlein}, K. and {Kolleck}, M. and {Lagg}, A. and {Meller}, R. and {M{\"u}ller}, R. and {Sch{\"u}hle}, U. and {Staub}, J. and {Albert}, K. and {Alvarez Copano}, M. and {Beckmann}, U. and {Bischoff}, J. and {Busse}, D. and {Enge}, R. and {Frahm}, S. and {Germerott}, D. and {Guerrero}, L. and {L{\"o}ptien}, B. and {Meierdierks}, T. and {Oberdorfer}, D. and {Papagiannaki}, I. and {Ramanath}, S. and {Schou}, J. and {Werner}, S. and {Yang}, D. and {Zerr}, A. and {Bergmann}, M. and {Bochmann}, J. and {Heinrichs}, J. and {Meyer}, S. and {Monecke}, M. and {M{\"u}ller}, M. -F. and {Sperling}, M. and {{\'A}lvarez Garc{\'\i}a}, D. and {Aparicio}, B. and {Balaguer Jim{\'e}nez}, M. and {Bellot Rubio}, L.~R. and {Cobos Carracosa}, J.~P. and {Girela}, F. and {Hern{\'a}ndez Exp{\'o}sito}, D. and {Herranz}, M. and {Labrousse}, P. and {L{\'o}pez Jim{\'e}nez}, A. and {Orozco Su{\'a}rez}, D. and {Ramos}, J.~L. and {Barandiar{\'a}n}, J. and {Bastide}, L. and {Campuzano}, C. and {Cebollero}, M. and {D{\'a}vila}, B. and {Fern{\'a}ndez-Medina}, A. and {Garc{\'\i}a Parejo}, P. and {Garranzo-Garc{\'\i}a}, D. and {Laguna}, H. and {Mart{\'\i}n}, J.~A. and {Navarro}, R. and {N{\'u}{\~n}ez Peral}, A. and {Royo}, M. and {S{\'a}nchez}, A. and {Silva-L{\'o}pez}, M. and {Vera}, I. and {Villanueva}, J. and {Fourmond}, J. -J. and {de Galarreta}, C. Ruiz and {Bouzit}, M. and {Hervier}, V. and {Le Clec'h}, J.~C. and {Szwec}, N. and {Chaigneau}, M. and {Buttice}, V. and {Dominguez-Tagle}, C. and {Philippon}, A. and {Boumier}, P. and {Le Cocguen}, R. and {Baranjuk}, G. and {Bell}, A. and {Berkefeld}, Th. and {Baumgartner}, J. and {Heidecke}, F. and {Maue}, T. and {Nakai}, E. and {Scheiffelen}, T. and {Sigwarth}, M. and {Soltau}, D. and {Volkmer}, R. and {Blanco Rodr{\'\i}guez}, J. and {Domingo}, V. and {Ferreres Sabater}, A. and {Gasent Blesa}, J.~L. and {Rodr{\'\i}guez Mart{\'\i}nez}, P. and {Osorno Caudel}, D. and {Bosch}, J. and {Casas}, A. and {Carmona}, M. and {Herms}, A. and {Roma}, D. and {Alonso}, G. and {G{\'o}mez-Sanjuan}, A. and {Piqueras}, J. and {Torralbo}, I. and {Fiethe}, B. and {Guan}, Y. and {Lange}, T. and {Michel}, H. and {Bonet}, J.~A. and {Fahmy}, S. and {M{\"u}ller}, D. and {Zouganelis}, I.},
        title = "{The Polarimetric and Helioseismic Imager on Solar Orbiter}",
      journal = {\aap},
         year = 2020,
        month = oct,
       volume = {642},
          eid = {A11},
        pages = {A11},
          doi = {10.1051/0004-6361/201935325},
archivePrefix = {arXiv},
       eprint = {1903.11061},
 primaryClass = {astro-ph.IM},
       adsurl = {https://ui.adsabs.harvard.edu/abs/2020A&A...642A..11S}
}

@ARTICLE{Solo,
       author = {{M{\"u}ller}, D. and {St. Cyr}, O.~C. and {Zouganelis}, I. and {Gilbert}, H.~R. and {Marsden}, R. and {Nieves-Chinchilla}, T. and {Antonucci}, E. and {Auch{\`e}re}, F. and {Berghmans}, D. and {Horbury}, T.~S. and {Howard}, R.~A. and {Krucker}, S. and {Maksimovic}, M. and {Owen}, C.~J. and {Rochus}, P. and {Rodriguez-Pacheco}, J. and {Romoli}, M. and {Solanki}, S.~K. and {Bruno}, R. and {Carlsson}, M. and {Fludra}, A. and {Harra}, L. and {Hassler}, D.~M. and {Livi}, S. and {Louarn}, P. and {Peter}, H. and {Sch{\"u}hle}, U. and {Teriaca}, L. and {del Toro Iniesta}, J.~C. and {Wimmer-Schweingruber}, R.~F. and {Marsch}, E. and {Velli}, M. and {De Groof}, A. and {Walsh}, A. and {Williams}, D.},
        title = "{The Solar Orbiter mission. Science overview}",
      journal = {\aap},
         year = 2020,
        month = oct,
       volume = {642},
          eid = {A1},
        pages = {A1},
          doi = {10.1051/0004-6361/202038467},
archivePrefix = {arXiv},
       eprint = {2009.00861},
 primaryClass = {astro-ph.SR},
       adsurl = {https://ui.adsabs.harvard.edu/abs/2020A&A...642A...1M}
}

@ARTICLE{Kim2019,
       author = {{Kim}, Taeyoung and {Park}, Eunsu and {Lee}, Harim and {Moon}, Yong-Jae and {Bae}, Sung-Ho and {Lim}, Daye and {Jang}, Soojeong and {Kim}, Lokwon and {Cho}, Il-Hyun and {Choi}, Myungjin and {Cho}, Kyung-Suk},
        title = "{Solar farside magnetograms from deep learning analysis of STEREO/EUVI data}",
      journal = {Nature Astronomy},
         year = 2019,
        month = mar,
       volume = {3},
        pages = {397-400},
          doi = {10.1038/s41550-019-0711-5},
       adsurl = {https://ui.adsabs.harvard.edu/abs/2019NatAs...3..397K}
}

@ARTICLE{Jeong2022,
       author = {{Jeong}, Hyun-Jin and {Moon}, Yong-Jae and {Park}, Eunsu and {Lee}, Harim and {Baek}, Ji-Hye},
        title = "{Improved AI-generated Solar Farside Magnetograms by STEREO and SDO Data Sets and Their Release}",
      journal = {\apjs},
         year = 2022,
        month = oct,
       volume = {262},
       number = {2},
          eid = {50},
        pages = {50},
          doi = {10.3847/1538-4365/ac8d66},
archivePrefix = {arXiv},
       eprint = {2204.12068},
 primaryClass = {astro-ph.SR},
       adsurl = {https://ui.adsabs.harvard.edu/abs/2022ApJS..262...50J}
}

@inproceedings{stanzione2020frontera,
author = {Stanzione, Dan and West, John and Evans, R. Todd and Minyard, Tommy and Ghattas, Omar and Panda, Dhabaleswar K.},
title = {Frontera: The Evolution of Leadership Computing at the National Science Foundation},
year = {2020},
isbn = {9781450366892},
publisher = {Association for Computing Machinery},
url = {https://doi.org/10.1145/3311790.3396656},
doi = {10.1145/3311790.3396656},
booktitle = {Practice and Experience in Advanced Research Computing 2020: Catch the Wave},
pages = {106–111},
numpages = {6}
}

@ARTICLE{2003SoPh..212..165S,
       author = {{Schrijver}, Carolus J. and {De Rosa}, Marc L.},
        title = "{Photospheric and heliospheric magnetic fields}",
      journal = {\solphys},
         year = 2003,
        month = jan,
       volume = {212},
       number = {1},
        pages = {165-200},
          doi = {10.1023/A:1022908504100},
       adsurl = {https://ui.adsabs.harvard.edu/abs/2003SoPh..212..165S}
}

@ARTICLE{2001ApJ...547..475S,
       author = {{Schrijver}, Carolus J.},
        title = "{Simulations of the Photospheric Magnetic Activity and Outer Atmospheric Radiative Losses of Cool Stars Based on Characteristics of the Solar Magnetic Field}",
      journal = {\apj},
         year = 2001,
        month = jan,
       volume = {547},
       number = {1},
        pages = {475-490},
          doi = {10.1086/318333},
       adsurl = {https://ui.adsabs.harvard.edu/abs/2001ApJ...547..475S}
}

@ARTICLE{1994Koutchmy,
       author = {{Koutchmy}, S.},
        title = "{Coronal physics from eclipse observations}",
      journal = {Advances in Space Research},
         year = 1994,
        month = apr,
       volume = {14},
       number = {4},
        pages = {29-39},
          doi = {10.1016/0273-1177(94)90156-2},
       adsurl = {https://ui.adsabs.harvard.edu/abs/1994AdSpR..14d..29K}
}

@ARTICLE{2014Druckmuller,
       author = {{Druckm{\"u}ller}, Miloslav and {Habbal}, Shadia Rifai and {Morgan}, Huw},
        title = "{Discovery of a New Class of Coronal Structures in White Light Eclipse Images}",
      journal = {\apj},
         year = 2014,
        month = apr,
       volume = {785},
       number = {1},
          eid = {14},
        pages = {14},
          doi = {10.1088/0004-637X/785/1/14},
       adsurl = {https://ui.adsabs.harvard.edu/abs/2014ApJ...785...14D}
}

@ARTICLE{1958ApJ...128..664P,
       author = {{Parker}, E.~N.},
        title = "{Dynamics of the Interplanetary Gas and Magnetic Fields.}",
      journal = {\apj},
         year = 1958,
        month = nov,
       volume = {128},
        pages = {664},
          doi = {10.1086/146579},
       adsurl = {https://ui.adsabs.harvard.edu/abs/1958ApJ...128..664P}
}

@ARTICLE{2024ZhuSpectr,
       author = {{Zhu}, Yingjie and {Habbal}, Shadia R. and {Ding}, Adalbert and {Yamashiro}, Bryan and {Landi}, Enrico and {Boe}, Benjamin and {Constantinou}, Sage and {Nassir}, Michael},
        title = "{Spectroscopic Observations of the Solar Corona during the 2017 August 21 Total Solar Eclipse: Comparison of Spectral Line Widths and Doppler Shifts between Open and Closed Magnetic Structures}",
      journal = {\apj},
         year = 2024,
        month = may,
       volume = {966},
       number = {1},
          eid = {122},
        pages = {122},
          doi = {10.3847/1538-4357/ad3424},
archivePrefix = {arXiv},
       eprint = {2403.10363},
 primaryClass = {astro-ph.SR},
       adsurl = {https://ui.adsabs.harvard.edu/abs/2024ApJ...966..122Z}
}

@article{boe2018first,
  title={The first empirical determination of the Fe10+ and Fe13+ freeze-in distances in the solar corona},
  author={Boe, Benjamin and Habbal, Shadia and Druckm{\"u}ller, Miloslav and Landi, Enrico and Kourkchi, Ehsan and Ding, Adalbert and Starha, Pavel and Hutton, Joseph},
  journal={\apj},
  volume={859},
  number={2},
  pages={155},
  year={2018},
  doi={10.3847/1538-4357/aabfb7},
  publisher={IOP Publishing}
}

@article{boe2021color,
  title={The color and brightness of the F-corona inferred from the 2019 July 2 total solar eclipse},
  author={Boe, Benjamin and Habbal, Shadia and Downs, Cooper and Druckm{\"u}ller, Miloslav},
  journal={\apj},
  volume={912},
  number={1},
  pages={44},
  year={2021},
  doi={10.3847/1538-4357/abea79},
  publisher={IOP Publishing}
}

@article{toth2005space,
  title={{Space Weather Modeling Framework: A New Tool for the Space Science Community}},
  author={T{\'o}th, G{\'a}bor and Sokolov, Igor V and Gombosi, Tamas I and Chesney, David R and Clauer, C Robert and De Zeeuw, Darren L and Hansen, Kenneth C and Kane, Kevin J and Manchester, Ward B and Oehmke, Robert C and others},
  journal={Journal of Geophysical Research: Space Physics},
  volume={110},
  number={A12},
  year={2005},
  doi={10.1029/2005JA011126},
  publisher={Wiley Online Library}
}

@article{toth2012adaptive,
  title={Adaptive numerical algorithms in space weather modeling},
  author={T{\'o}th, G{\'a}bor and van der Holst, Bart and Sokolov, Igor V and De Zeeuw, Darren L and Gombosi, Tamas I and Fang, Fang and Manchester, Ward B and Meng, Xing and Najib, Dalal and Powell, Kenneth G and others},
  journal={Journal of Computational Physics},
  volume={231},
  number={3},
  pages={870--903},
  year={2012},
  doi={10.1016/j.jcp.2011.02.006},
  publisher={Elsevier}
}

@article{sokolov2013mhd,
  title={{Magnetohydrodynamic Waves and Coronal Heating: Unifying Empirical and MHD Turbulence Models}},
  author={Sokolov, Igor V and van der Holst, Bart and Oran, Rona and Downs, Cooper and Roussev, Ilia I and Jin, Meng and Manchester, Ward B and Evans, Rebekah M and Gombosi, Tamas I},
  journal={\apj},
  volume={764},
  number={1},
  pages={23},
  year={2013},
  doi={10.1088/0004-637X/764/1/23},
  publisher={IOP Publishing}
}

@article{gombosi2003adaptive,
  title={Adaptive mesh refinement for global magnetohydrodynamic simulation},
  author={Gombosi, Tamas I and De Zeeuw, Darren L and Powell, Kenneth G and Ridley, Aaron J and Sokolov, Igor V and Stout, Quentin F and T{\'o}th, G{\'a}bor},
  journal={Space Plasma Simulation},
  pages={247--274},
  year={2003},
  doi={10.1007/3-540-36530-3_12},
  publisher={Springer}
}

@ARTICLE{Milic2024,
       author = {{Mili{\'c}}, I. and {Centeno}, R. and {Sun}, X. and {Rempel}, M. and {de la Cruz Rodr{\'\i}guez}, J.},
        title = "{Spatial resolution effects on the solar open flux estimates}",
      journal = {\aap},
         year = 2024,
        month = mar,
       volume = {683},
          eid = {A134},
        pages = {A134},
          doi = {10.1051/0004-6361/202347272},
archivePrefix = {arXiv},
       eprint = {2402.02486},
 primaryClass = {astro-ph.SR},
       adsurl = {https://ui.adsabs.harvard.edu/abs/2024A&A...683A.134M}
}

@ARTICLE{Krivova2004,
       author = {{Krivova}, N.~A. and {Solanki}, S.~K.},
        title = "{Effect of spatial resolution on estimating the Sun's magnetic flux}",
      journal = {\aap},
         year = 2004,
        month = apr,
       volume = {417},
        pages = {1125-1132},
          doi = {10.1051/0004-6361:20040022},
       adsurl = {https://ui.adsabs.harvard.edu/abs/2004A&A...417.1125K}
}

@ARTICLE{BelloRubio2019,
       author = {{Bellot Rubio}, Luis and {Orozco Su{\'a}rez}, David},
        title = "{Quiet Sun magnetic fields: an observational view}",
      journal = {Living Reviews in Solar Physics},
         year = 2019,
        month = dec,
       volume = {16},
       number = {1},
          eid = {1},
        pages = {1},
          doi = {10.1007/s41116-018-0017-1},
       adsurl = {https://ui.adsabs.harvard.edu/abs/2019LRSP...16....1B}
}

@ARTICLE{Ruvsin2010,
       author = {{Ru{\v{s}}in}, V. and {Druckm{\"u}ller}, M. and {Aniol}, P. and {Minarovjech}, M. and {Saniga}, M. and {Miki{\'c}}, Z. and {Linker}, J.~A. and {Lionello}, R. and {Riley}, P. and {Titov}, V.~S.},
        title = "{Comparing eclipse observations of the 2008 August 1 solar corona with an MHD model prediction}",
      journal = {\aap},
         year = 2010,
        month = apr,
       volume = {513},
          eid = {A45},
        pages = {A45},
          doi = {10.1051/0004-6361/200912778},
       adsurl = {https://ui.adsabs.harvard.edu/abs/2010A&A...513A..45R}
}

@ARTICLE{Parnell2012,
       author = {{Parnell}, C.~E. and {De Moortel}, I.},
        title = "{A contemporary view of coronal heating}",
      journal = {Philosophical Transactions of the Royal Society of London Series A},
         year = 2012,
        month = jul,
       volume = {370},
       number = {1970},
        pages = {3217-3240},
          doi = {10.1098/rsta.2012.0113},
archivePrefix = {arXiv},
       eprint = {1206.6097},
 primaryClass = {astro-ph.SR},
       adsurl = {https://ui.adsabs.harvard.edu/abs/2012RSPTA.370.3217P}
}

@ARTICLE{Velli2015,
       author = {{Velli}, M. and {Pucci}, F. and {Rappazzo}, F. and {Tenerani}, A.},
        title = "{Models of coronal heating, turbulence and fast reconnection}",
      journal = {Philosophical Transactions of the Royal Society of London Series A},
         year = 2015,
        month = apr,
       volume = {373},
       number = {2042},
        pages = {20140262-20140262},
          doi = {10.1098/rsta.2014.0262},
       adsurl = {https://ui.adsabs.harvard.edu/abs/2015RSPTA.37340262V}
}

@ARTICLE{Downs2016,
       author = {{Downs}, Cooper and {Lionello}, Roberto and {Miki{\'c}}, Zoran and {Linker}, Jon A. and {Velli}, Marco},
        title = "{Closed-field Coronal Heating Driven by Wave Turbulence}",
      journal = {\apj},
         year = 2016,
        month = dec,
       volume = {832},
       number = {2},
          eid = {180},
        pages = {180},
          doi = {10.3847/0004-637X/832/2/180},
archivePrefix = {arXiv},
       eprint = {1610.02113},
 primaryClass = {astro-ph.SR},
       adsurl = {https://ui.adsabs.harvard.edu/abs/2016ApJ...832..180D}
}

@ARTICLE{Downs2025,
       author = {{Downs}, Cooper and {Linker}, Jon A. and {Caplan}, Ronald M. and {Mason}, Emily I. and {Riley}, Pete and {Davidson}, Ryder and {Reyes}, Andres and {Palmerio}, Erika and {Lionello}, Roberto and {Turtle}, James and {Ben-Nun}, Michal and {Stulajter}, Miko M. and {Titov}, Viacheslav S. and {T{\"o}r{\"o}k}, Tibor and {Upton}, Lisa A. and {Attie}, Raphael and {Jha}, Bibhuti K. and {Arge}, Charles N. and {Henney}, Carl J. and {Valori}, Gherardo and {Strecker}, Hanna and {Calchetti}, Daniele and {Germerott}, Dietmar and {Hirzberger}, Johann and {Su{\'a}rez}, David Orozco and {Rodr{\'\i}guez}, Julian Blanco and {Solanki}, Sami K. and {Cheng}, Xin and {Wu}, Sizhe},
        title = "{A near-real-time data-assimilative model of the solar corona}",
      journal = {Science},
         year = 2025,
        month = jun,
       volume = {388},
       number = {6753},
        pages = {1306-1310},
          doi = {10.1126/science.adq0872},
       adsurl = {https://ui.adsabs.harvard.edu/abs/2025Sci...388.1306D}
}

@ARTICLE{Shi2022,
       author = {{Shi}, Tong and {Manchester}, IV, Ward and {Landi}, Enrico and {van der Holst}, Bart and {Szente}, Judit and {Chen}, Yuxi and {T{\'o}th}, G{\'a}bor and {Bertello}, Luca and {Pevtsov}, Alexander},
        title = "{AWSoM Magnetohydrodynamic Simulation of a Solar Active Region with Realistic Spectral Synthesis}",
      journal = {\apj},
         year = 2022,
        month = mar,
       volume = {928},
       number = {1},
          eid = {34},
        pages = {34},
          doi = {10.3847/1538-4357/ac52ab},
       adsurl = {https://ui.adsabs.harvard.edu/abs/2022ApJ...928...34S}
}

@ARTICLE{Chitta2023,
       author = {{Chitta}, L.~P. and {Seaton}, D.~B. and {Downs}, C. and {DeForest}, C.~E. and {Higginson}, A.~K.},
        title = "{Direct observations of a complex coronal web driving highly structured slow solar wind}",
      journal = {Nature Astronomy},
         year = 2023,
        month = feb,
       volume = {7},
        pages = {133-141},
          doi = {10.1038/s41550-022-01834-5},
archivePrefix = {arXiv},
       eprint = {2211.13283},
 primaryClass = {astro-ph.SR},
       adsurl = {https://ui.adsabs.harvard.edu/abs/2023NatAs...7..133C}
}

@ARTICLE{Riley2006,
       author = {{Riley}, Pete and {Linker}, J.~A. and {Miki{\'c}}, Z. and {Lionello}, R. and {Ledvina}, S.~A. and {Luhmann}, J.~G.},
        title = "{A Comparison between Global Solar Magnetohydrodynamic and Potential Field Source Surface Model Results}",
      journal = {\apj},
         year = 2006,
        month = dec,
       volume = {653},
       number = {2},
        pages = {1510-1516},
          doi = {10.1086/508565},
       adsurl = {https://ui.adsabs.harvard.edu/abs/2006ApJ...653.1510R}
}

@ARTICLE{Gressl2014,
       author = {{Gressl}, C. and {Veronig}, A.~M. and {Temmer}, M. and {Odstr{\v{c}}il}, D. and {Linker}, J.~A. and {Miki{\'c}}, Z. and {Riley}, P.},
        title = "{Comparative Study of MHD Modeling of the Background Solar Wind}",
      journal = {\solphys},
         year = 2014,
        month = may,
       volume = {289},
       number = {5},
        pages = {1783-1801},
          doi = {10.1007/s11207-013-0421-6},
archivePrefix = {arXiv},
       eprint = {1312.1220},
 primaryClass = {astro-ph.SR},
       adsurl = {https://ui.adsabs.harvard.edu/abs/2014SoPh..289.1783G}
}

@ARTICLE{Henadhira2022,
       author = {{Henadhira Arachchige}, Kalpa and {Cohen}, Ofer and {Munoz-Jaramillo}, Andres and {Yeates}, Anthony R.},
        title = "{Comparing the Performance of a Solar Wind Model from the Sun to 1 au Using Real and Synthetic Magnetograms}",
      journal = {\apj},
         year = 2022,
        month = oct,
       volume = {938},
       number = {1},
          eid = {39},
        pages = {39},
          doi = {10.3847/1538-4357/ac8d59},
archivePrefix = {arXiv},
       eprint = {2208.13668},
 primaryClass = {astro-ph.SR},
       adsurl = {https://ui.adsabs.harvard.edu/abs/2022ApJ...938...39H}
}

@ARTICLE{Linker2017,
       author = {{Linker}, J.~A. and {Caplan}, R.~M. and {Downs}, C. and {Riley}, P. and {Mikic}, Z. and {Lionello}, R. and {Henney}, C.~J. and {Arge}, C.~N. and {Liu}, Y. and {Derosa}, M.~L. and {Yeates}, A. and {Owens}, M.~J.},
        title = "{The Open Flux Problem}",
      journal = {\apj},
         year = 2017,
        month = oct,
       volume = {848},
       number = {1},
          eid = {70},
        pages = {70},
          doi = {10.3847/1538-4357/aa8a70},
archivePrefix = {arXiv},
       eprint = {1708.02342},
 primaryClass = {astro-ph.SR},
       adsurl = {https://ui.adsabs.harvard.edu/abs/2017ApJ...848...70L}
}

@ARTICLE{Shi2025,
       author = {{Shi}, Guanglu and {Shan}, Jiahui and {Feng}, Li and {Chen}, Jun and {Gan}, Weiqun},
        title = "{The Role of Far-side Magnetic Structures in Modeling 2024 Solar Eclipse}",
      journal = {arXiv e-prints},
         year = 2025,
        month = sep,
          eid = {arXiv:2509.02911},
        pages = {arXiv:2509.02911},
          doi = {10.48550/arXiv.2509.02911},
archivePrefix = {arXiv},
       eprint = {2509.02911},
 primaryClass = {astro-ph.SR},
       adsurl = {https://ui.adsabs.harvard.edu/abs/2025arXiv250902911S}
}

@ARTICLE{Shi2024,
       author = {{Shi}, Tong and {Manchester}, Ward and {Landi}, Enrico and {van der Holst}, Bart and {Szente}, Judit and {Chen}, Yuxi and {T{\'o}th}, G{\'a}bor and {Bertello}, Luca and {Pevtsov}, Alexander},
        title = "{AWSoM Magnetohydrodynamic Simulation of a Solar Active Region. II. Statistical Analysis of Alfv{\'e}n Wave Dissipation and Reflection, Scaling Laws, and Energy Budget on Coronal Loops}",
      journal = {\apj},
         year = 2024,
        month = jan,
       volume = {961},
       number = {1},
          eid = {60},
        pages = {60},
          doi = {10.3847/1538-4357/ad0df2},
       adsurl = {https://ui.adsabs.harvard.edu/abs/2024ApJ...961...60S}
}

@ARTICLE{Caplan2025,
       author = {{Caplan}, Ronald M. and {Stulajter}, Miko M. and {Linker}, Jon A. and {Downs}, Cooper and {Upton}, Lisa A. and {Jha}, Bibhuti Kumar and {Attie}, Raphael and {Arge}, Charles N. and {Henney}, Carl J.},
        title = "{Open-source Flux Transport (OFT). I. HipFT{\textendash}High-performance Flux Transport}",
      journal = {\apjs},
         year = 2025,
        month = may,
       volume = {278},
       number = {1},
          eid = {24},
        pages = {24},
          doi = {10.3847/1538-4365/adc080},
archivePrefix = {arXiv},
       eprint = {2501.06377},
 primaryClass = {astro-ph.SR},
       adsurl = {https://ui.adsabs.harvard.edu/abs/2025ApJS..278...24C}
}

@ARTICLE{Del-Zanna2015,
       author = {{Del Zanna}, G. and {Dere}, K.~P. and {Young}, P.~R. and {Landi}, E. and {Mason}, H.~E.},
        title = "{CHIANTI - An atomic database for emission lines. Version 8}",
      journal = {\aap},
         year = 2015,
        month = oct,
       volume = {582},
          eid = {A56},
        pages = {A56},
          doi = {10.1051/0004-6361/201526827},
archivePrefix = {arXiv},
       eprint = {1508.07631},
 primaryClass = {astro-ph.SR},
       adsurl = {https://ui.adsabs.harvard.edu/abs/2015A&A...582A..56D}
}

@ARTICLE{Krieger1973,
       author = {{Krieger}, A.~S. and {Timothy}, A.~F. and {Roelof}, E.~C.},
        title = "{A Coronal Hole and Its Identification as the Source of a High Velocity Solar Wind Stream}",
      journal = {\solphys},
         year = 1973,
        month = apr,
       volume = {29},
       number = {2},
        pages = {505-525},
          doi = {10.1007/BF00150828},
       adsurl = {https://ui.adsabs.harvard.edu/abs/1973SoPh...29..505K}
}

@INPROCEEDINGS{Bohlin1977,
       author = {{Bohlin}, J.~D.},
        title = "{An observational definition of coronal holes.}",
    booktitle = {Coronal Holes and High Speed Wind Streams},
         year = 1977,
       editor = {{Zirker}, J.~B.},
        month = jan,
        pages = {27-69},
       adsurl = {https://ui.adsabs.harvard.edu/abs/1977chhs.conf...27B}
}

@ARTICLE{Cranmer2009,
       author = {{Cranmer}, Steven R.},
        title = "{Coronal Holes}",
      journal = {Living Reviews in Solar Physics},
         year = 2009,
        month = dec,
       volume = {6},
       number = {1},
          eid = {3},
        pages = {3},
          doi = {10.12942/lrsp-2009-3},
archivePrefix = {arXiv},
       eprint = {0909.2847},
 primaryClass = {astro-ph.SR},
       adsurl = {https://ui.adsabs.harvard.edu/abs/2009LRSP....6....3C}
}

@ARTICLE{Feldman1992,
       author = {{Feldman}, Uri},
        title = "{REVIEW:  Elemental abundances in the upper solar atmosphere}",
      journal = {\physscr},
         year = 1992,
        month = sep,
       volume = {46},
       number = {3},
        pages = {202-220},
          doi = {10.1088/0031-8949/46/3/002},
       adsurl = {https://ui.adsabs.harvard.edu/abs/1992PhyS...46..202F}
}

@INPROCEEDINGS{VigilESA,
       author = {{Palomba}, Massimo and {Luntama}, Juha-Pekka},
        title = "{Vigil: ESA Space Weather Mission in L5}",
    booktitle = {44th COSPAR Scientific Assembly. Held 16-24 July},
         year = 2022,
       volume = {44},
        month = jul,
        pages = {3544},
       adsurl = {https://ui.adsabs.harvard.edu/abs/2022cosp...44.3544P}
}

@ARTICLE{Feng2015,
       author = {{Feng}, Xueshang and {Ma}, Xiaopeng and {Xiang}, Changqing},
        title = "{Data-driven modeling of the solar wind from 1 R$_{s}$ to 1 AU}",
      journal = {Journal of Geophysical Research (Space Physics)},
         year = 2015,
        month = dec,
       volume = {120},
       number = {12},
        pages = {10,159-10,174},
          doi = {10.1002/2015JA021911},
       adsurl = {https://ui.adsabs.harvard.edu/abs/2015JGRA..12010159F}
}

@misc{Pevtsov2022_NSOHMI,
  author       = {{Pevtsov}, Alexander and {Bertello}, Luca},
  title        = {NSO-HMI-NRT Synoptic Maps of Vector Magnetic Field},
  year         = {2022},
  publisher    = {WIM Research \& Education},
  doi          = {10.25668/nw0t-b078},
  url          = {https://doi.org/10.25668/nw0t-b078}
}

@dataset{derosa_2025_lm_fasam_hmi,
  author       = {DeRosa, M.},
  title        = {LM-FASAM-HMI Forecast Magnetic Maps For 2024 TSE},
  year         = {2025},
  publisher    = {Zenodo},
  doi          = {10.5281/zenodo.17459827},
  url          = {https://doi.org/10.5281/zenodo.17459827}
}

@ARTICLE{Manchester2004,
       author = {{Manchester}, Ward B. and {Gombosi}, Tamas I. and {Roussev}, Ilia and {de Zeeuw}, Darren L. and {Sokolov}, I.~V. and {Powell}, Kenneth G. and {T{\'o}th}, G{\'a}Bor and {Opher}, Merav},
        title = "{Three-dimensional MHD simulation of a flux rope driven CME}",
      journal = {Journal of Geophysical Research (Space Physics)},
         year = 2004,
        month = jan,
       volume = {109},
       number = {A1},
          eid = {A01102},
        pages = {A01102},
          doi = {10.1029/2002JA009672},
       adsurl = {https://ui.adsabs.harvard.edu/abs/2004JGRA..109.1102M}
}

@string{AA			=	"Astron. Astrophys."}

@string{ACADEMIC		=	"Academic Press"}

@string{AG			=	"Ann. Geophys."}

@string{APJ			=	"Astrophys. J."}

@string{APJL			=	"Astrophys. J. Lett."}

@string{APJS			=	"Astrophys. J. Suppl."}

@string{JCP			=	"J. Comput. Phys."}

@string{NATURE			=	"Nature"}

@string{NM			=	"Numer. Math."}

@string{PP			=	"Phys. Plasmas"}

@string{SCI			=	"Science"}

@string{SOLPHYS			=	"Sol. Phys."}

@string{SPRINGER		=	"Springer-Verlag"}

@string{WILEY			=	"John Wiley \& Sons"}

@string{apjs = "{Astro.~Phys.~J.~Supp.}"}

@string{aap = "{A\&A}"}

@string{nature = "{Nature}"}

@string{pp = "{Phys.~Plasmas}"}

@book{Billings:1966,
  author = "Billings, D. W.",
  title = "A guide to the  solar corona",
  publisher = "Academic", 
  address = "San Diego, Calif.", 
  year = 1966}

@article{Chen:2016,
  author = "Y. Chen and G. T\'oth and T. I. Gombosi",
  title = "A fifth-order finite difference scheme for hyperbolic equations on block-adaptive curvilinear grids",
  journal=JCP,
  volume=305,
  pages = 604,
  year = 2016,
  doi = "10.1016/j.jcp.2015.11.003"}

@ARTICLE{Hoeksema:2014,
 author = {{Hoeksema}, J.~T. and {Liu}, Y. and {Hayashi}, K. and {Sun}, X. and 
    {Schou}, J. and {Couvidat}, S. and {Norton}, A. and {Bobra}, M. and 
    {Centeno}, R. and {Leka}, K.~D. and {Barnes}, G. and {Turmon}, M.
    },
  title = "{The Helioseismic and Magnetic Imager (HMI) Vector Magnetic 
            Field Pipeline: Overview and Performance}",
  Journal = SOLPHYS,
     year = 2014,
   volume = 289,
    pages = {3483-3530},
      doi = {10.1007/s11207-014-0516-8}
}

@inproceedings{Koren:1993,
  author="Koren, B.", 
  title="A robust upwind discretisation method for advection, 
         diffusion and source terms",
  booktitle="Numerical Methods for Advection-Diffusion Problems",
  editor = "C.B. Vreugdenhil and B.Koren",
  publisher = "Vieweg, Braunschweig",
  pages = 117,
  year = 1993}

@article{Mikic:1999,
                  AUTHOR = {Z. Mikic and J.A. Linker and D.D. Schnack 
                   and R. Lionello and A. Tarditi},
                  TITLE = {Magnetohydrodynamic modeling of the global 
                           solar corona},
                  YEAR = {1999},
                  JOURNAL = PP,
                  VOLUME = {6},
                  PAGES = {2217-2224}}

@article{Suresh:1997,
  author = "A. Suresh and H. T. Huynh",
  title = "Accurate monotonicity-preserving schemes with 
           Runge-Kutta time stepping",
  journal = JCP,
  volume = 136,
  pages = 83,
  year = 1997}

@ARTICLE{Sachdeva:2021,
       author = {{Sachdeva}, N. and {T{\'o}th}, G. and {Manchester}, W. B. and {van der Holst}, B. and {Huang}, Z. and {Sokolov}, I. V. and {Zhao}, L. and {Shidi}, Q. A. and {Chen}, Y.and {Gombosi}, T. I. and {Henney}, C. J. and {Lloveras}, D. G. and {V{\'a}squez}, A. M.},
        title = "{Simulating Solar Maximum Conditions Using the Alfv{\'e}n Wave Solar Atmosphere Model (AWSoM)}",
      journal = {\apj},
         year = 2021,
        month = dec,
       volume = {923},
       number = {2},
          eid = {176},
        pages = {176},
          doi = {10.3847/1538-4357/ac307c},
       adsurl = {https://ui.adsabs.harvard.edu/abs/2021ApJ...923..176S}
}

@ARTICLE{Sachdeva:2019,
       author = {{Sachdeva}, N. and {van der Holst}, B. and {Manchester}, W. B. and {T{\'o}th}, G. and {Chen}, Y.and {Lloveras}, D. G. and {V{\'a}squez}, A. M. and {Lamy}, P. and {Wojak}, J. and {Jackson}, B. V. and {Yu}, H.-S. and {Henney}, C. J.},
        title = "{Validation of the Alfv{\'e}n Wave Solar Atmosphere Model (AWSoM) with Observations from the Low Corona to 1 au}",
      journal = {\apj},
         year = 2019,
        month = dec,
       volume = {887},
       number = {1},
          eid = {83},
        pages = {83},
          doi = {10.3847/1538-4357/ab4f5e},
archivePrefix = {arXiv},
       eprint = {1910.08110},
 primaryClass = {astro-ph.SR},
       adsurl = {https://ui.adsabs.harvard.edu/abs/2019ApJ...887...83S}
}

@ARTICLE{Bertello:2014,
       author = {{Bertello}, L. and {Pevtsov}, A.~A. and {Petrie}, G.~J.~D. and {Keys}, D.},
        title = "{Uncertainties in Solar Synoptic Magnetic Flux Maps}",
      journal = {\solphys},
         year = 2014,
        month = jul,
       volume = {289},
       number = {7},
        pages = {2419-2431},
          doi = {10.1007/s11207-014-0480-3},
archivePrefix = {arXiv},
       eprint = {1312.0509},
 primaryClass = {astro-ph.SR},
       adsurl = {https://ui.adsabs.harvard.edu/abs/2014SoPh..289.2419B}
}

@ARTICLE{Usmanov:2018,
       author = {{Usmanov}, Arcadi V. and {Matthaeus}, William H. and {Goldstein}, Melvyn L. and {Chhiber}, Rohit},
        title = "{The Steady Global Corona and Solar Wind: A Three-dimensional MHD Simulation with Turbulence Transport and Heating}",
      journal = {\apj},
         year = 2018,
        month = sep,
       volume = {865},
       number = {1},
          eid = {25},
        pages = {25},
          doi = {10.3847/1538-4357/aad687},
       adsurl = {https://ui.adsabs.harvard.edu/abs/2018ApJ...865...25U}
}

@ARTICLE{Huang:2023,
       author = {{Huang}, Zhenguang and {T{\'o}th}, G{\'a}bor and {Sachdeva}, Nishtha and {Zhao}, Lulu and {van der Holst}, Bart and {Sokolov}, Igor and {Manchester}, Ward B. and {Gombosi}, Tamas I.},
        title = "{Modeling the Solar Wind during Different Phases of the Last Solar Cycle}",
      journal = {\apjl},
         year = 2023,
        month = apr,
       volume = {946},
       number = {2},
          eid = {L47},
        pages = {L47},
          doi = {10.3847/2041-8213/acc5ef},
       adsurl = {https://ui.adsabs.harvard.edu/abs/2023ApJ...946L..47H}
}

@ARTICLE{Sachdeva:2023,
       author = {{Sachdeva}, Nishtha and {Manchester}, Ward B., IV and {Sokolov}, Igor and {Huang}, Zhenguang and {Pevtsov}, Alexander and {Bertello}, Luca and {Pevtsov}, Alexei A. and {Toth}, Gabor and {van der Holst}, Bart and {Henney}, Carl J.},
        title = "{Solar Wind Modeling with the Alfv{\'e}n Wave Solar atmosphere Model Driven by HMI-based Near-real-time Maps by the National Solar Observatory}",
      journal = {\apj},
         year = 2023,
        month = aug,
       volume = {952},
       number = {2},
          eid = {117},
        pages = {117},
          doi = {10.3847/1538-4357/acda87},
archivePrefix = {arXiv},
       eprint = {2212.05138},
 primaryClass = {astro-ph.SR},
       adsurl = {https://ui.adsabs.harvard.edu/abs/2023ApJ...952..117S}
}

@ARTICLE{Huang:2024GONG,
       author = {{Huang}, Zhenguang and {T{\'o}th}, G{\'a}bor and {Sachdeva}, Nishtha and {van der Holst}, Bart},
        title = "{Solar Wind Driven from GONG Magnetograms in the Last Solar Cycle}",
      journal = {\apj},
         year = 2024,
        month = apr,
       volume = {965},
       number = {1},
          eid = {1},
        pages = {1},
          doi = {10.3847/1538-4357/ad32ca},
archivePrefix = {arXiv},
       eprint = {2403.01656},
 primaryClass = {astro-ph.SR},
       adsurl = {https://ui.adsabs.harvard.edu/abs/2024ApJ...965....1H}
}
\bibliographystyle{aasjournal}

\end{document}